\documentclass[useAMS,usenatbib]{mnras}

\pdfoutput=1 

\usepackage[english]{babel}
\usepackage{amsmath,amssymb}
\usepackage{graphicx, subfig}
\usepackage[normalem]{ulem}

\usepackage{verbatim}

\usepackage{color}
\usepackage{multirow}
\usepackage{mathtools}
\usepackage{epstopdf}

\usepackage{url}
\usepackage{color}

\def\nat{Nature}

\def\mnras{MNRAS}
\def\apj{ApJ}
\def\apjl{ApJL}
\def\apjs{ApJS}
\def\aap{A\&A}
\def\aaps{A\&A Supp.}

\def\araa{ARA\&A}

\def\pasp{Publ. Astr. Soc. Pac.}

\def\be{\begin{equation}} 
\def\ee{\end{equation}} 
\def\ba{\begin{eqnarray}} 
\def\ea{\end{eqnarray}}

\def\cc{\,{\rm {cm^{-3}}}} 
\def\msun{{\Msun}}

\def\HH{${\rm {H_2}}$}

\def\HII{\hbox{H~$\scriptstyle\rm II\ $}}

\def\gsim{\lower.5ex\hbox{\gtsima}} 
\def\lsim{\lower.5ex\hbox{\ltsima}} \def\gtsima{$\; \buildrel > \over 
\sim \;$} \def\ltsima{$\; \buildrel < \over \sim \;$} \def\prosima{$\; 
\buildrel \propto \over \sim \;$} \def\gsim{\lower.5ex\hbox{\gtsima}} 
\def\lsim{\lower.5ex\hbox{\ltsima}} 
\def\simgt{\lower.5ex\hbox{\gtsima}} 
\def\simlt{\lower.5ex\hbox{\ltsima}} 
\def\simpr{\lower.5ex\hbox{\prosima}} \def\la{\lsim} \def\ga{\gsim} 
  
 \def\gtsima{$\; \buildrel > \over \sim \;$} 
\def\ltsima{$\; \buildrel < \over \sim \;$} 
\def\gsim{\lower.5ex\hbox{\gtsima}} 
\def\lsim{\lower.5ex\hbox{\ltsima}} 
\def\simgt{\lower.5ex\hbox{\gtsima}} 
\def\simlt{\lower.5ex\hbox{\ltsima}} 
\def\simpr{\lower.5ex\hbox{\prosima}} 
\def\la{\lsim} 
\def\ga{\gsim}

\def\msun{\,{\rm \Msun}}

\def\E3{{\cal E}_{\rm g}^{III}}

\def\r12{r_{1/2}} 
\def\x12{x_{1/2}} 
\def\v12{v_{1/2}}

\def\HII{\hbox{H~$\scriptstyle\rm II $~}}

\def\CII{\hbox{[C~$\scriptstyle\rm II $]}}
\def\CIIion{\hbox{C~$\scriptstyle\rm II $}}

\def\OIII{\hbox{[O~$\scriptstyle\rm III $]}}

\newcommand\textlcsc[1]{\textsc{\MakeLowercase{#1}}}
\newcommand{\quotes}[1]{``#1''}

\def\angstrom{\textrm{A\kern -1.3ex\raisebox{0.6ex}{$^\circ$}}}
\def\myr{\rm Myr}

\def\msun{{\rm M}_{\odot}}
\def\zsun{{\rm Z}_{\odot}}
\def\lsun{{\rm L}_{\odot}}

\def\dust{\mathcal{D}}
\def\dsun{\dust_{\odot}}

\def\nh2{n_{\rm H2}}
\def\fh2{f_{\rm H2}}

\def\msunyr{\msun\,{\rm yr}^{-1}}
\def\cc{{\rm cm}^{-3}}

\def\surfl{\lsun\,{\rm kpc}^{-2}}

\defcitealias{krumholz:2009apj}{KTM09}
\defcitealias{pallottini:2017dahlia}{P17}
\defcitealias{bovino:2016aa}{B16}
\defcitealias{vallini:2015}{V15}
\defcitealias{jiang:2016apj}{J16}

\def\KTMcit{\citetalias{krumholz:2009apj}}
\def\P17cit{\citetalias{pallottini:2017dahlia}}
\def\B16cit{\citetalias{bovino:2016aa}}
\def\V15cit{\citetalias{vallini:2015}}
\def\J16cit{\citetalias{jiang:2016apj}}

\def\highz{$\mbox{high-}z$~}

\begin{document}

\date{}
\pagerange{\pageref{firstpage}--\pageref{lastpage}} \pubyear{2017}
\title[Chemistry in \highz galaxies]{The impact of chemistry on the structure of \highz galaxies}
\author[Pallottini et al.]{A. Pallottini$^{1,2,3,4}$\thanks{\href{mailto:andrea.pallottini@centrofermi.it}{andrea.pallottini@centrofermi.it}; \href{mailto:ap926@mrao.cam.ac.uk}{ap926@mrao.cam.ac.uk}},
 A. Ferrara$^{4,5}$,
 S. Bovino$^{6}$,
 L. Vallini$^{7}$,
 S. Gallerani$^{4}$,\newauthor
 R. Maiolino$^{2,3}$,
 S. Salvadori$^{8,9,10}$.\\\\
$^{1}$ Centro Fermi, Museo Storico della Fisica e Centro Studi e Ricerche ``Enrico Fermi'', Piazza del Viminale 1, Roma, 00184, Italy\\
$^{2}$ Cavendish Laboratory, University of Cambridge, 19 J. J. Thomson Ave., Cambridge CB3 0HE, UK\\
$^{3}$ Kavli Institute for Cosmology, University of Cambridge, Madingley Road, Cambridge CB3 0HA, UK\\
$^{4}$ Scuola Normale Superiore, Piazza dei Cavalieri 7, I-56126 Pisa, Italy\\
$^{5}$ Kavli IPMU, The University of Tokyo, 5-1-5 Kashiwanoha, Kashiwa 277-8583, Japan\\
$^{6}$ Hamburger Sternwarte, Universit{\"a}t Hamburg, Gojenbergsweg 112, D-21029 Hamburg, Germany\\
$^{7}$ Nordita, KTH Royal Institute of Technology and Stockholm University, Roslagstullsbacken 23, SE-10691 Stockholm, Sweden\\
$^{8}$ Dipartimento di Fisica e Astronomia, Universita' di Firenze, Via G. Sansone 1, Sesto Fiorentino, Italy\\
$^{9}$ INAF/Osservatorio Astrofisico di Arcetri, Largo E. Fermi 5, Firenze, Italy\\
$^{10}$ GEPI, Observatoire de Paris, PSL Research University, CNRS, Place Jule Janssen 92190, Meudon, France
}

\maketitle

\label{firstpage}

\begin{abstract}
To improve our understanding of \highz galaxies we study the impact of \HH~chemistry on their evolution, morphology and observed properties. We compare two zoom-in high-resolution ($30\,{\rm pc}$) simulations of prototypical $M_{\star}\sim 10^{10}\msun$ galaxies at $z=6$. The first, \quotes{Dahlia}, adopts an equilibrium model for \HH~formation, while the second, \quotes{Alth{\ae}a}, features an improved non-equilibrium chemistry network.
The star formation rate (SFR) of the two galaxies is similar (within 50\%), and increases with time reaching values close to $100\, \msunyr$ at $z=6$. They both have SFR-stellar mass relation consistent with observations, and a specific SFR of $\simeq 5\, {\rm Gyr}^{-1}$.
The main differences arise in the gas properties. The non-equilibrium chemistry determines the H$\rightarrow$ \HH~transition to occur at densities $> 300\, \cc$, i.e. about 10 times larger than predicted by the equilibrium model used for Dahlia. As a result, Alth{\ae}a features a more clumpy and fragmented morphology, in turn making SN feedback more effective. Also, because of the lower density and weaker feedback, Dahlia sits $3\sigma$ away from the Schmidt-Kennicutt relation; Alth{\ae}a, instead nicely agrees with observations.
The different gas properties result in widely different observables. Alth{\ae}a outshines Dahlia by a factor of $7$ ($15$) in \CII~$157.74\,\mu{\rm m}$ (\HH~$17.03\,\mu{\rm m}$) line emission. Yet, Alth{\ae}a is under-luminous with respect to the locally observed \CII-SFR relation. Whether this relation does not apply at \highz or the line luminosity is reduced by CMB and metallicity effects remains as an open question. 

\end{abstract}

\begin{keywords}
galaxies: high-redshift, formation, evolution, ISM -- infrared: general -- methods: numerical
\end{keywords}

%----------------------------------------------------------------------------------------
\section{Introduction}
%----------------------------------------------------------------------------------------

Understanding the properties of the interstellar medium (ISM) of primeval galaxies is a fundamental challenge of physical cosmology.
The high sensitivity/spatial resolution allowed by current observations have dramatically improved our understanding of the ISM of local and moderate redshift ($z=2-3$) galaxies \citep{osterbrock:1989book,stasinska:2007,perezmontero:2017,stanway:2017}. We now have a clearer picture of the gas phases and thermodynamics \citep{daddi:2010apj,carilli:2013ara&a}, particularly for what concerns the molecular component, representing the stellar birth environment \citep{klessen:2014review,krumholz:2015review}.

For galaxies located in the Epoch of Reionization (EoR, $5 \lsim z \lsim 15$) optical/near infrared (IR) surveys have been very successful in their identification and characterization in terms of stellar mass and star formation rate \citep{Dunlop13,Madau14,Bouwens:2015}. However, only recently we have started to probe the internal structure of such objects. With the advent of the Atacama Large Millimeter/Submillmeter Array (ALMA) it is now possible to access the far infrared (FIR) band at \highz with an unprecedented resolution and sensitivity. Excitingly, this enables for the first time studies of ISM energetics, structure and composition in such pristine objects.

Since \CIIion~is one of the major coolant of the ISM, \CII~ALMA detections (and upper limits) have so far mostly used this line for the above purposes \citep{maiolino:2015arxiv,willott:2015arxiv15,capak:2015arxiv} and to determine the sizes of early galaxies \citep{fujimoto:2017}. Line emission from different species (e.g. \OIII) have been used to derive the interstellar radiation field (ISRF) intensity \citep[][]{inoue:2016sci,carniani:2017bdf3299}, while continuum detections give us a measure of the dust content and properties \citep[][]{watson:2015nature,laporte:2017apj}. Finally, some observations are beginning to resolve different ISM components and their dynamics by detecting spatial offsets and kinematic shifts between different emission lines, i.e. \CII~and optical-ultraviolet (UV) emission \citep{maiolino:2015arxiv,capak:2015arxiv}, \CII~and Ly$\alpha$ \citep{pentericci:2016apj,bradac:2017} and \CII~and \OIII~\citep{carniani:2017bdf3299}.

In spite of these progresses, several pressing questions remain unanswered. A partial list includes: {(a)} What is the chemical composition and thermodynamic state of the ISM in \highz galaxies? {(b)} How does the molecular gas turns into stars and regulate the evolution of these systems? {(c)} What are the optimal observational strategies to better constrain the properties of these primeval objects?

Theoretically, cosmological numerical simulations have been used to attack some of these problems. The key idea is to produce a coherent physical framework within which the observed properties can be understood. Such learning strategy is also of fundamental importance to devise efficient observations from current (e.g. HST/ALMA), planned (JWST) and proposed (SPICA) instruments.
Before this strategy can be implemented, though, it is necessary to develop reliable numerical schemes catching all the relevant physical processes. While the overall performances of the most widely used schemes have been extensively benchmarked \citep{aquila:2012MNRAS,agora:2013arxiv,kim:2016apj}, high-resolution simulations of galaxy formation introduce a new challenge: they are very sensitive to the implemented physical models, particularly those acting on small scales.

Among these, the role of feedback, i.e. how stars affect their own formation history via energy injection in the surrounding gas by supernova (SN) explosions, stellar winds and radiation, is far from being completely understood, despite considerable efforts to improve its modeling \citep{agertz:2015apj,martizzi:2015mnras} and understand its consequences on \highz galaxy evolution \citep{ceverino:2014,oshea:2015apj,barai:2015mnras,pallottini:2017dahlia,fiacconi:2017mnras,hopkins:2017}.

Additionally, we are still lacking a completely self-consistent treatment of radiation transfer. This is an area in which intensive work is ongoing in terms of faster numerical schemes \citep{wise:2012radpres,rosdahl:2015mnras,katz:2016arxiv}, or improved physical modelling \citep{petkova:2012mnras,roskar:2014,maio:2016mnras}.

A third aspect has received comparatively less attention so far in \highz galaxy formation studies, i.e. the implementation of adequate chemical networks. While various models have been proposed and tested \citep{krumholz:2009apj,bovino:2016aa,grassi:2017dust}, the galaxy-scale consequences of the different prescriptions are still largely unexplored \citep{tomassetti:2015MNRAS,maio:2015,smith:2017mnras}. Besides, there is no clear consensus on a minimal set of physical ingredients required to produce reliable simulations.

The purpose of this paper is to analyze the impact of \HH~chemistry on the internal structure of \highz galaxies. To this aim, we simulate two prototypical $M_{\star}\simeq 10^{10}\msun$ Lyman Break Galaxies (LBG) at $z=6$, named \quotes{Dahlia} and \quotes{Alth{\ae}a}, respectively. The two simulations differ for the \HH~formation implementation, equilibrium vs. non-equilibrium. We show how chemistry has a strong impact on the observed properties of early galaxies.

The paper is organized as follows. In Sec.~\ref{sec_numerical} we describe the two simulations highlighting common features (Sec.~\ref{sec_common_pre}), separately discussing the different chemical models used for Dahlia (Sec.~\ref{sec_chem_eq}) and Alth{\ae}a (Sec.~\ref{sec_chem_noneq}).
Results are presented as follow. First we perform a benchmark of the chemical models (Sec.~\ref{sec_chem_test}), and compare the star formation and feedback history of the two galaxies (Sec.~\ref{sec_sfr_feed}). Next, we characterize their differences in terms of morphology (Sec.~\ref{sec_morphology}), thermodynamical state of the ISM (Sec.~\ref{sec_thermo_state}), and predicted \CII~and \HH~(Sec.~\ref{sec_obs_prop}) emission line properties. Our conclusions are summarized in Sec.~\ref{sec_conclusione}.

%----------------------------------------------------------------------------------------
\section{Numerical simulations}\label{sec_numerical}
%----------------------------------------------------------------------------------------

To assess the impact of \HH~chemistry on the internal structure of \highz galaxies, we compare two zoom-in simulations adopting different chemical models. Both simulations follow the evolution of a prototypical $z=6$ LBG galaxy hosted by a $M_{\rm h}\simeq 10^{11} \msun$ dark matter (DM) halo (virial radius $r_{\rm vir}\simeq 15 \,{\rm kpc}$). 

The first simulation has been presented in \citet[][hereafter \P17cit]{pallottini:2017dahlia}. The targeted galaxy (which includes also about 10 satellites) is called \quotes{Dahlia} \citep[see also][for analysis of its infall/outflow structure]{gallerani:2016outflow}. In such previous work we showed that Dahlia's specific SFR (sSFR) is in agreement with both analytical calculations \citep{behroozi:2013apj}, and with $z=7$ observations \citep[][see also Sec.~\ref{sec_sfr_feed}]{gonzalez:2010}.

In the second, new simulation we follow the evolution of \quotes{Alth{\ae}a}, by using improved thermo-chemistry, but keeping everything else (initial conditions, resolution, star formation and feedback prescriptions) unchanged with respect to the Dahlia simulation. We describe the implementation of these common processes in the following Section. Next we describe separately the chemical model used for Dahlia (Sec.~\ref{sec_chem_eq}) and Alth{\ae}a (Sec.~\ref{sec_chem_noneq}).

\subsection{Common physical models}\label{sec_common_pre}
Both simulations are performed with a customized version of the Adaptive Mesh Refinement (AMR) code \textlcsc{ramses} \citep{teyssier:2002}. Starting from cosmological IC\footnote{We assume cosmological parameters compatible with \emph{Planck} results: $\Lambda$CDM model with total matter, vacuum and baryonic densities in units of the critical density $\Omega_{\Lambda}= 0.692$, $\Omega_{m}= 0.308$, $\Omega_{b}= 0.0481$, Hubble constant $\rm H_0=100\,{\rm h}\,{\rm km}\,{\rm s}^{-1}\,{\rm Mpc}^{-1}$ with ${\rm h}=0.678$, spectral index $n=0.967$, $\sigma_{8}=0.826$ \citep[][]{planck:2013_xvi_parameters}.} generated with \textlcsc{music} \citep{hahn:2011mnras}, we zoom-in the $z\simeq 6$ DM halo hosting the targeted galaxy.
The total simulation volume is $(20\,{\rm Mpc}/{\rm h})^{3}$ that is evolved with a base grid with 8 levels (gas mass $6\times 10^6\msun$); the zoom-in region has a volume of $(2.1\,{\rm Mpc}/{\rm h})^{3}$ and is resolved with 3 additional level of refinement, thus yielding a gas mass resolution of $m_b = 1.2\times 10^4 \msun$. In such region, we allow for 6 additional level of refinement, that allow to follow the evolution of the gas down to scales of $l_{\rm cell}\simeq 30\,{\rm pc}$ at $z=6$, i.e. the refined cells have mass and size typical of Galactic molecular clouds \citep[MC, e.g.][]{federrath:2013}. The refinement is performed with a Lagrangian mass threshold-based criterion.
, i.e. a cell is refined if its total (DM+baryonic) mass exceed the the mass resolution by a factor 8.

Metallicity ($Z$) is followed as the sum of heavy elements, assumed to have solar abundance ratios \citep{asplund:2009ara&a}. We impose an initial metallicity floor $Z_{\rm floor}=10^{-3}\zsun$ since at $z \gsim 40$ our resolution is still insufficient to catch the metal enrichment by the first stars \citep[e.g.][]{oshea:2015apj}. Such floor is compatible with the metallicity found at \highz in cosmological simulations for diffuse enriched gas \citep{dave:2011mnras,pallottini:2014sim,maio:2015}; it only marginally affects the gas cooling time.

Dust evolution is not explicitly tracked during simulations. However, we make the simple assumption that the dust-to-gas mass ratio scales with metallicity, i.e. $\dust = \dsun (Z/\zsun)$, where $\dsun/\zsun = 0.3$ for the Milky Way (MW, e.g. \citealt[][]{hirashita:2002mnras,asano:2013}.

\subsubsection{Star formation}

Stars form according to a linearly \HH-dependent Schmidt-Kennicutt relation \citep[][]{schmidt:1959apj,kennicutt:1998apj} i.e.
\be\label{eq_sk_relation}
\dot{\rho}_{\star}= \zeta_{\rm sf} \fh2 {\rho \over t_{\rm ff}},
\ee
where $\dot{\rho}_{\star}$ is the local SF rate density, $\zeta_{\rm sf}$ the SF efficiency, $\fh2$ the \HH~mass fraction, and $\rho =\mu m_p n$ is density of the gas of mean molecular weight $\mu$.
Eq. \ref{eq_sk_relation} is solved stochastically, by drawing the mass of the new star particles from a Poisson distribution \citep{rasera:2006,dubois:2008,pallottini:2014sim}.
In detail, in a star formation event we create a star particle with mass $Nm_b$, with $N$ an integer drawn from
\be
P(N) = {\langle N\rangle \over N!} \exp-\langle N\rangle\,,
\ee
where the mean of the Poisson distribution is
\be
\langle N\rangle =  {\fh2 \rho l_{\rm cell}^{3}\over m_b} { \zeta_{\rm sf} \delta t\over t_{\rm ff}}\,,
\ee
with $\delta t$ the simulation time step. For numerical stability, no more than half of the cell mass is allowed to turn into stars. Since we prevent formation of star particle with mass less then $m_b$, cells with density less then $\sim 15\,\cc$ (for $l_{\rm cell}\simeq 30\,\rm pc$) are not allowed to form stars.

We set $\zeta_{\rm sf}=0.1$, in accordance with the average values inferred from MC observations \citep[][see also \citealt{agertz:2012arxiv}]{murray:2011apj}; $\fh2$ depends on the adopted thermo-chemical model, as described later in Sec.~\ref{sec_chem_eq} and Sec.~\ref{sec_chem_noneq}.

\subsubsection{Feedback}
Similarly to \citet[][]{agora:2013arxiv}, we account for stellar energy inputs and chemical yields that depend both on time and stellar populations by using \textlcsc{starburst99} \citep{starburst99:1999}. Stellar tracks are taken from the {\tt padova} \citep{padova:1994} library with stellar metallicities in the range $0.02 \leq Z_{\star}/\zsun \leq 1$, and we assume a \citet{kroupa:2001} initial mass function.
Stellar feedback includes SNs, winds from massive stars and radiation pressure \citep[][]{agertz:2012arxiv}. We model the thermal and turbulent energy content of the gas according to the prescriptions by \citet{agertz:2015apj}.
The turbulent (or non-thermal) energy is dissipated as $\dot{e}_{\rm nth} = -e_{\rm nth}/t_{\rm diss}$ \citep[][see eq. 2]{teyssier:2013mnras}, where, following \citet{maclow1999turb}, the dissipation time scale can be written as 
\be
t_{\rm diss} = 9.785 \left(l_{\rm cell}\over 100\,{\rm pc}\right)\left(\sigma_{\rm turb}\over 10\,{\rm km}\,{\rm s}^{-1}\right) ^{-1}\,\myr\,,
\ee
where $\sigma_{\rm turb}$ is the turbulent velocity dispersion. Adopting the SN blastwave models and OB/AGB stellar winds from \citet{ostriker:1988rvmp} and \citet{weaver:1977apj}, respectively, we account for the dissipation of energy in MCs as detailed in Sec.~2.4 and App. A of \P17cit.
%
% Chemistry
%

\subsection{Dahlia: equilibrium thermo-chemistry}\label{sec_chem_eq}

In the Dahlia simulation we compute $\fh2$ by adopting the \KTMcit~analytical prescription \citep{krumholz:2008apj,krumholz:2009apj,mckee:2010apj}. In \KTMcit, the \HH~abundance is derived by modelling the radiative transfer on an idealized MC and by assuming equilibrium between \HH~formation on dust grains and dissociation rates. For each gas cell, $\fh2$ can then be written as a function of $n$, $Z$ and hydrogen column density ($N_{\rm H}$). By further assuming pressure equilibrium between CNM and WNM \citep{krumholz:2009apj}, $\fh2$ turns out to be independent on the intensity of the ISRF, and can be written as
\begin{subequations}\label{eq_fh2_anal}
\begin{align}
\fh2 &= \left[1 -0.75\,s/(1+0.25\,s) \right]\Theta(2-s)\,,\\ \intertext{with}
s    &= \ln\left(1+0.6\,\chi +0.01\chi^{2}\right) /(0.6\,\tau_{\rm UV})\\
\chi &= 0.75\,\left[1+3.1\,(Z/\zsun)^{0.365}\right]\,,
\end{align}
\end{subequations}
and where $\Theta$ is the Heaviside function; $\tau_{\rm UV}$ is the dust UV optical depth and it can be calculated by linearly rescaling the MW value,
\be
\tau_{\rm UV} = \left({N_{\rm H}\over 1.6 \times 10^{21}{\rm cm}^{-2}}\right)\left({\dust\over\dsun}\right).
\ee

In Dahlia cooling/heating rates are computed using \textlcsc{grackle} 2.1\footnote{\url{https://grackle.readthedocs.org/}} \citep{bryan:2014apjs}. We use a \hbox{H}~and \hbox{He}~primordial network, and tabulated metal cooling/photo-heating rates from \textlcsc{cloudy} \citep{cloudy:2013}. Inverse Compton cooling is also present, and we consider heating from a redshift-dependent ionizing UV background \citep[UVB, ][]{Haardt:2012}. Since \HH~is not explicitly included in the network, we do not include the corresponding cooling contribution.

\subsection{Alth{\ae}a: non-equilibrium thermo-chemistry}\label{sec_chem_noneq}

In Alth{\ae}a we implement a non-equilibrium chemical network by using \textlcsc{krome}\footnote{\url{https://bitbucket.org/tgrassi/krome}} \citep{grassi:2014mnras}. Given a set of species and their reactions, \textlcsc{krome} can generate the code needed to solve the system of coupled ordinary differential equations that describe the gas thermo-chemical evolution.

\subsubsection{Chemical network}

Similarly to \citet[][hereafter \B16cit]{bovino:2016aa}, our network includes H, H$^{+}$, H$^{-}$, He, He$^{+}$, He$^{++}$, H$_{2}$, H$_{2}^{+}$ and electrons. Metal species are not followed individually in the network, as for instance done in Model IV from \B16cit; therefore, we use an equilibrium metal line cooling calculated via \textlcsc{cloudy} tables\footnote{As a caveat, we point out that there is a formal inconsistency in the modelling. Metal line cooling tables are usually calculated with \textlcsc{cloudy} by assuming a \citet{Haardt:2012} UV background, while the ISRF spectral energy density we adopt is MW-like. To remove such inconsistency one should explicitly track metal species, adopt a non-equilibrium metal line cooling and include radiative transfer. As noted in \B16cit (see their Fig.~16), using non-equilibrium metal line cooling can typically change the cooling function by a factor $\lsim 2$. This will be addressed in future work.}.
The adopted network contains a total of 37 reactions, including photo-chemistry (Sec.~\ref{sec_sub_photo}), dust processes (Sec.~\ref{sec_sub_dust}) and cosmic rays (CR, Sec.~\ref{sec_sub_cr}). The reactions, their rates, and corresponding references are listed in App. B of \B16cit: specifically we use reactions from 1 to 31 (Tab. B.1 in \B16cit), 53, 54, and from 58 to 61 (Tab. B.2 in \B16cit).

\subsubsection{Photo-chemistry}\label{sec_sub_photo}
Photo-chemistry cross sections are taken from \citet{verner:1996apjs} and by using the SWRI\footnote{\url{http://phidrates.space.swri.edu}.} and Leiden\footnote{\url{http://home.strw.leidenuniv.nl/~ewine/photo/}.} databases.
In the present simulation, the ISRF is not evolved self-consistently and it is approximated as follows.
For the spectral energy density (SED), we assume a MW like spectrum \citep{black:1987,draine:1978apjs}, and we specify the SED using 10 energy bins from $0.75$ eV to $14.16$ eV. Beyond 13.6 eV the flux drops to zero, i.e. we do not include ionizing radiation.

We consider a spatially uniform ISRF whose intensity is rescaled with the SFR such that $G = G_{0} ({\rm SFR}/\msunyr)$, where $G_{0}=1.6\times 10^{-3} {\rm erg}\,{\rm cm}^{-2}\,{\rm s}^{-1}$ is the far UV (FUV) flux in the Habing band ($6-13.6\,{\rm eV}$) normalized to the average MW value \citep{habing:1968}. Because of their sub-kpc sizes (\citealt{shibuya:2015apjs}, \citealt{fujimoto:2017}) high $G_0$ values are expected in typical LBG at $z\simeq 6$, as inferred also by \citet{carniani:2017bdf3299}. A similar situation is seen in some local dwarf galaxies \citep{cormier:2015} that are generally considered as local counterparts of \highz galaxies.
It is worth noting that the spatial variation of $G$ is very small in the MW, with an r.m.s. value $\simeq 3\,G_0$ \citep{habing:1968,wolfire:2003apj}. Nonetheless, spatial fluctuations of the ISRF, if present, might play some role in 
the evolution of \highz galaxies \citep[e.g.][]{katz:2016arxiv}. We will analyze this effect in future work.

On top of the ISRF, we consider the cosmic microwave background (CMB), that effectively sets a temperature floor for the gas. Additionally, we neglect the cosmic UVB, since the typical ISM densities are sufficiently large to ensure an efficient self-shielding \citep[e.g.][]{gnedin:2010}. For example, \citet{rahmati:2013mnras} have shown that at $z\simeq 5$ the hydrogen ionization due to the UVB is negligible for $n \gsim 10^{-2}\cc$, the typical density of diffuse ISM.
The self-shielding of \HH~to photo-dissociation is accounted by using the \citet{richings:2014} prescription\footnote{The self-shielding formulation by \citet{richings:2014} does not account for a directional dependence as done in more computationally costly models \citep{hartwig:2015apj}.}, thus in each gas cell the shielding can be expressed as an analytical function of its \HH~column density, temperature and turbulence \citep[cfr. with][]{wolcottgreen:2011}.

\subsubsection{Dust processes}\label{sec_sub_dust}
As for Dahlia, the dust mass is proportional to the metal mass. Here we also specify the dust size distribution to be the one appropriate for \highz galaxies, the Small Magellanic Cloud one, following \citet{weingartner:2001apj}.
Dust grains can affect the chemistry through cooling\footnote{Dust cooling is not included in the current model, as it gives only a minor contribution for $n<10^4\cc$, i.e. see Fig.~3 in \B16cit.} \citep{hollenbach1979apjs}, photoelectric-heating \citep{bakes:1994apj}, and by mediating the formation of molecules \citep{cazaux2009aa}.
In particular, the formation rate of \HH~on dust grains is approximated following \citet{Jura:1975apj} 
\be\label{eq_jura}
R_{\rm H2-dust} = 3\times 10^{-17}n\,n_{\rm H} (\dust/\dsun)\,\cc\,{\rm s}^{-1}\,,
\ee
where $n_{\rm H}$ is the hydrogen density. Note that for $\dust\gsim 10^{-2}\dsun$ this dust channel is dominant with respect to gas-phase formation (e.g. reactions 6--7 and 9--10 B.1 in \B16cit).

\subsubsection{Cosmic rays}\label{sec_sub_cr}
CR ionization can become important in regions shielded from radiation, like MC interiors. We assume a CR hydrogen ionization rate $\propto$ SFR \citep{valle:2002apj} and normalized to the MW value \citep{webber:1998apj}:
\be
\zeta_{\rm cr} = 3\times 10^{-17} ({\rm SFR}/\msunyr)\, {\rm s}^{-1}.
\ee
The rate $\zeta_{\rm cr}$ includes the flux of CR and secondary electrons \citep[][]{richings:2014}. In the network, CR ionizations are proportional to $\zeta_{\rm cr}$ and to coupling constants that depend on the specific ions; such couplings are taken from the \textlcsc{kida} database \citep{kida:2012apjs}.
Additionally we account for Coulomb heating, by assuming that every CR ionization releases an energy\footnote{For a more accurate treatment of Coulomb heating refer to \citet[][]{glassgold:2012apj}.} of $20$ eV.

\subsubsection{Initial abundances of the species}

Finally, following \citet{galli:1998AA}, we calculate IC for the various species by accounting for the primordial chemistry\footnote{For a possible implementation of the \citet{galli:1998AA} chemical network see the \quotes{earlyUniverse} test contained in \textlcsc{krome}.} at $z\gsim 100$, for a density and temperature evolution corresponding to gas at the mean cosmic density.

\subsection{Benchmark of \HH~formation models}\label{sec_chem_test}

As a benchmark for our simulations, we compare the formation of \HH~in different physical environments. For the Dahlia \KTMcit~model we compute $\fh2$ from eq. \ref{eq_fh2_anal} as a function of $n$ and $Z$. We choose an expression for $N_{\rm H} = n\,l_{\rm cell}\mu\propto n^{2/3}$ resulting from the mass threshold-based AMR refinement criterion for which $l_{\rm cell} \propto n^{-1/3}$. We restate that the equilibrium \KTMcit~model is independent on $G$ and the gas temperature $T$.

For the Alth{\ae}a \B16cit~model we use \textlcsc{krome} to perform single-zone tests varying $n, Z$ and $G$. In this case we assume an initial temperature\footnote{The initial temperature corresponds to the virial temperature of the first star-forming halos present in the zoomed region. The results depend very weakly on this assumption.} $T= 5\times 10^3 {\rm K}$, and we let the gas patch evolve at constant density until thermo-chemical equilibrium is reached. This typically takes 100 Myr. 

\begin{figure}
\centering
\includegraphics[width=0.49\textwidth]{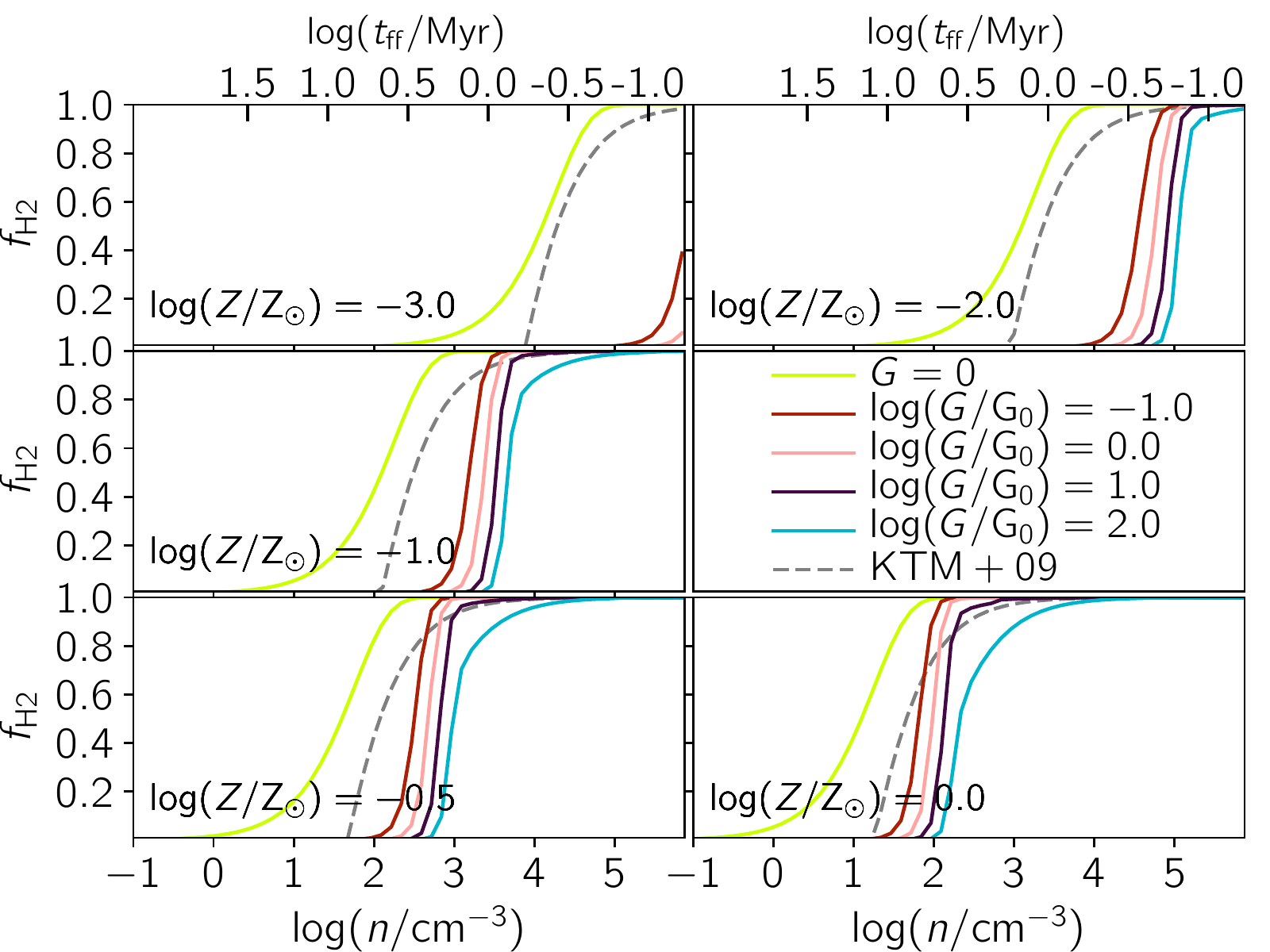}
\caption{Benchmark of the formation of \HH~for the model used in Dahlia (\KTMcit, Sec.~\ref{sec_chem_eq}) and in Alth{\ae}a (\B16cit, Sec.~\ref{sec_chem_noneq}).
In each panel we plot the \HH~mass fraction $\fh2$ as a function of density ($n$), with different panels showing the results for different metallicities ($Z$). In each panels the dashed grey line indicates the \KTMcit~model, while the \B16cit~models are plotted with solid lines, with different colours indicating a different impinging ISRF flux ($G$). In the upper axis we indicate the free-fall times ($t_{\rm ff}$) corresponding to $n$.
\label{fig_chimica_krome}
}
\end{figure}

The comparison between the two models is shown in Fig.~\ref{fig_chimica_krome} as a function of $n$ for different metallicities. 
For $G>0$ and $Z<\zsun$, \HH~formation is hindered in \B16cit~with respect to \KTMcit, i.e. higher $n$ are needed to reach similar $\fh2$ fractions. For $G = G_{0}$ and $Z=\zsun$ the two models are roughly in agreement: this is expected since \KTMcit~is calibrated on the MW environment. Finally, for $G>0$ and at $Z=10^{-3}\zsun$ (the metallicity floor in our simulation set) the \HH~formation in the \B16cit~model is strongly suppressed, e.g. $\fh2\simeq 10^{-3}$ for $n\simeq 10^3\cc$. Note that these fractions are comparable to the ones expected for \HH~formation in a pristine environment where \HH~formation proceeds via gas-phase reactions.

As noted in \P17cit, Dahlia's star formation (SF) model (eqs \ref{eq_sk_relation} and \ref{eq_fh2_anal}) is roughly equivalent to a density threshold criterion with metallicity-dependent critical density $n_{c} \simeq 26.45 \, (Z/Z_\odot)^{-0.87} \cc$. Physically this corresponds to the density at which $\fh2 \geq 0.5$ (see also \citealt{agertz:2012arxiv}). Thus, Fig.~\ref{fig_chimica_krome} quantifies the density threshold required to spawn stars in the simulation.

Dahlia forms stars in gas with $n\simeq 30\,\cc$ and $Z\simeq 0.5\,\zsun$ at a rate of about $10^2\msunyr$ at $z=6$. If Alth{\ae}a has a similar SFR history (this is checked a posteriori, see Fig.~\ref{fig_sfr_comparison}), the resulting metallicity and ISRF intensity ($G\simeq 10^2 G_{0}$) should also be similar. Then, by inspecting Fig.~\ref{fig_chimica_krome} (middle-left panel) one can conclude that Alth{\ae}a forms stars in much denser environments where $n > n_{c} \simeq 263 \, (Z/Z_\odot)^{-1.19} \cc$ for $G= 10^2 G_{0}$.
As noted in \citet{hopkins:2013arxiv}, although variations in the density threshold lead to similar total SFR, they might severely affect the galaxy morphology. We will return to this point in the next Section.

We remind that in both simulation we use the cell radius to calculate column density, that are used e.g. to calculate the gas self-shielding. This is done to mainly to ensure a fair comparison between the two simulations. 
In other simulations, e.g. MC illuminated by an external radiation field, the prescriptions adopted accounts for the contribution of column density from nearby cells, i.e. by using Jeans or Sobolev-like length (see e.g. \citealt{hartwig:2015apj} for a comparison between different prescriptions). However in our simulation we expect stars to be very close or embedded in potential star forming regions. Using the contribution to the column density from the surrounding gas would then overestimate the self-shielding effect. Such modelling uncertainty would be solved by including radiative transfer in the simulation.
However, we note that at z=6 the radius of our cells as a function of density can be approximated as $r_{\rm cell} = 154.1\,(n/\cc)^{-1/3}{\rm pc}$, while the jeans length is $l_J = 15.6\, (n/\cc)^{-1/2} (T/{\rm K})^{1/2}{\rm  pc}$. Thus for typical values found for the molecular gas in Althaea ($n \simeq 300 \,\cc$ and $T\simeq 100 \,\rm K$, see later Fig. \ref{fig_eos_h2}), the two prescriptions gives similar results, i.e. $r_{\rm cell} \sim 20 \,\rm pc$ and $l_J \sim 10\,\rm pc$.

\section{Results}

We now turn to a detailed analysis of the two zoomed galaxies, Dahlia and Alth{\ae}a\footnote{We refrain from the analysis of the satellite population of the two galaxies due to the oversimplifying assumption of a spatially uniform ISRF artificially suppressing star formation in environments with metallicity close to the floor value $Z_{\rm floor}=10^{-3}\zsun$.}. We start by studying the star formation and the build-up of the stellar mass from $z \simeq 15$ to $z=6$ (Sec.~\ref{sec_sfr_feed}). We then specialize at $z=6$ to inspect the galaxy morphology (Sec.~\ref{sec_morphology}), the ISM multiphase structure (Sec.~\ref{sec_thermo_state}) and the predicted observable properties (Sec.~\ref{sec_obs_prop}). An overall summary of the properties of the two galaxies is given in Tab.~\ref{tab_summary}. 
\subsection{Star formation history}\label{sec_sfr_feed}

\begin{figure}
\centering
\includegraphics[width=0.49\textwidth]{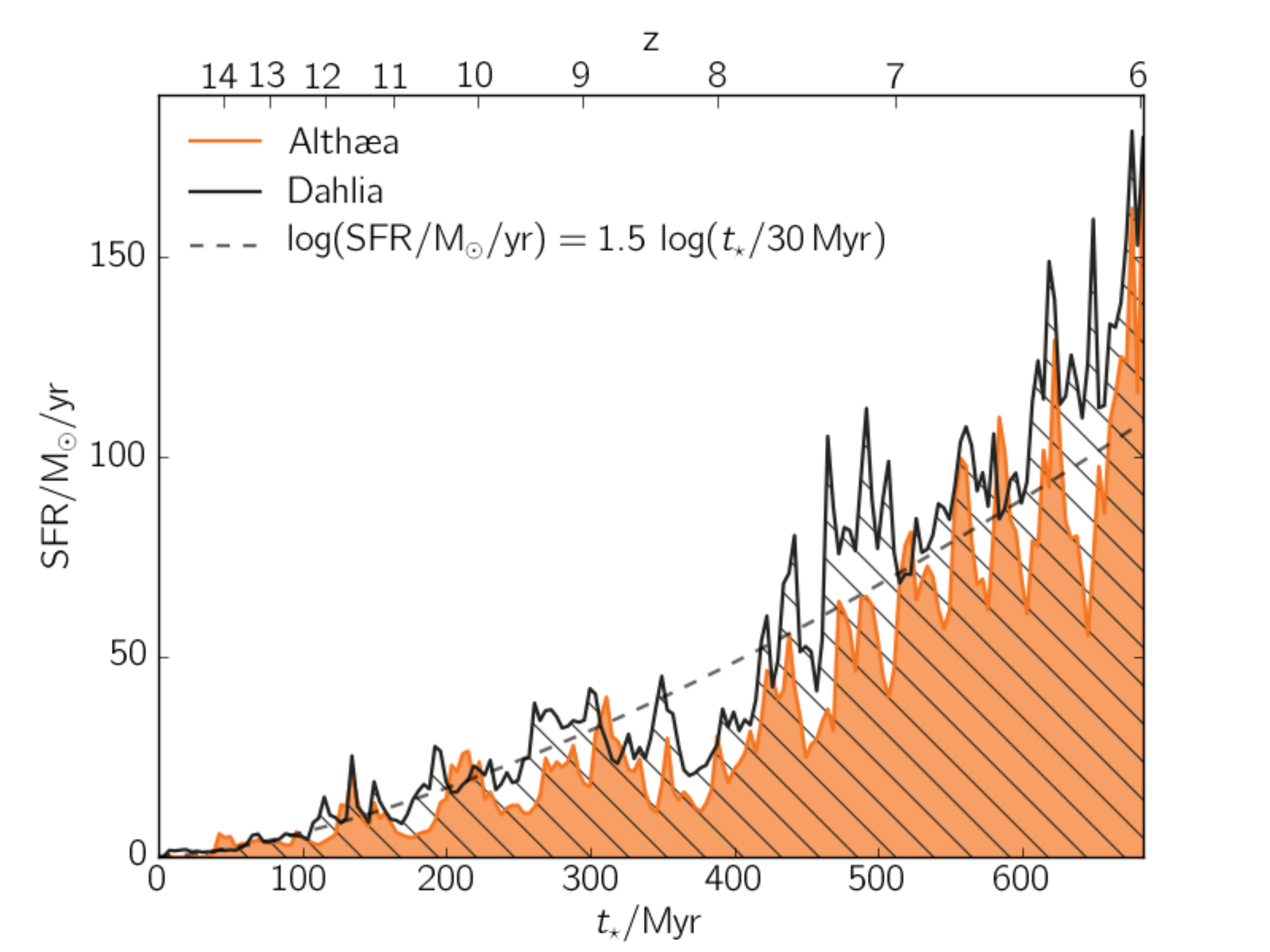}
\caption{
Star formation rate (SFR) as a function of galaxy age ($t_{\star}$) for Dahlia (black line and hatched region) and Alth{\ae}a (orange line and transparent region).
Also shown (grey dashed line) is an analytical approximation (within a factor 2 for both galaxies) to the average SFR trend. 
The redshift ($z$) corresponding to $t_{\star}$ is plotted on the upper axis, and note that $t_{\star}=0$ corresponds to the first stellar formation event in Dahlia, and the plotted SFRs are averaged over $4\,\myr$.
\label{fig_sfr_comparison}
}
\end{figure}

\begin{figure}
\centering
\includegraphics[width=0.49\textwidth]{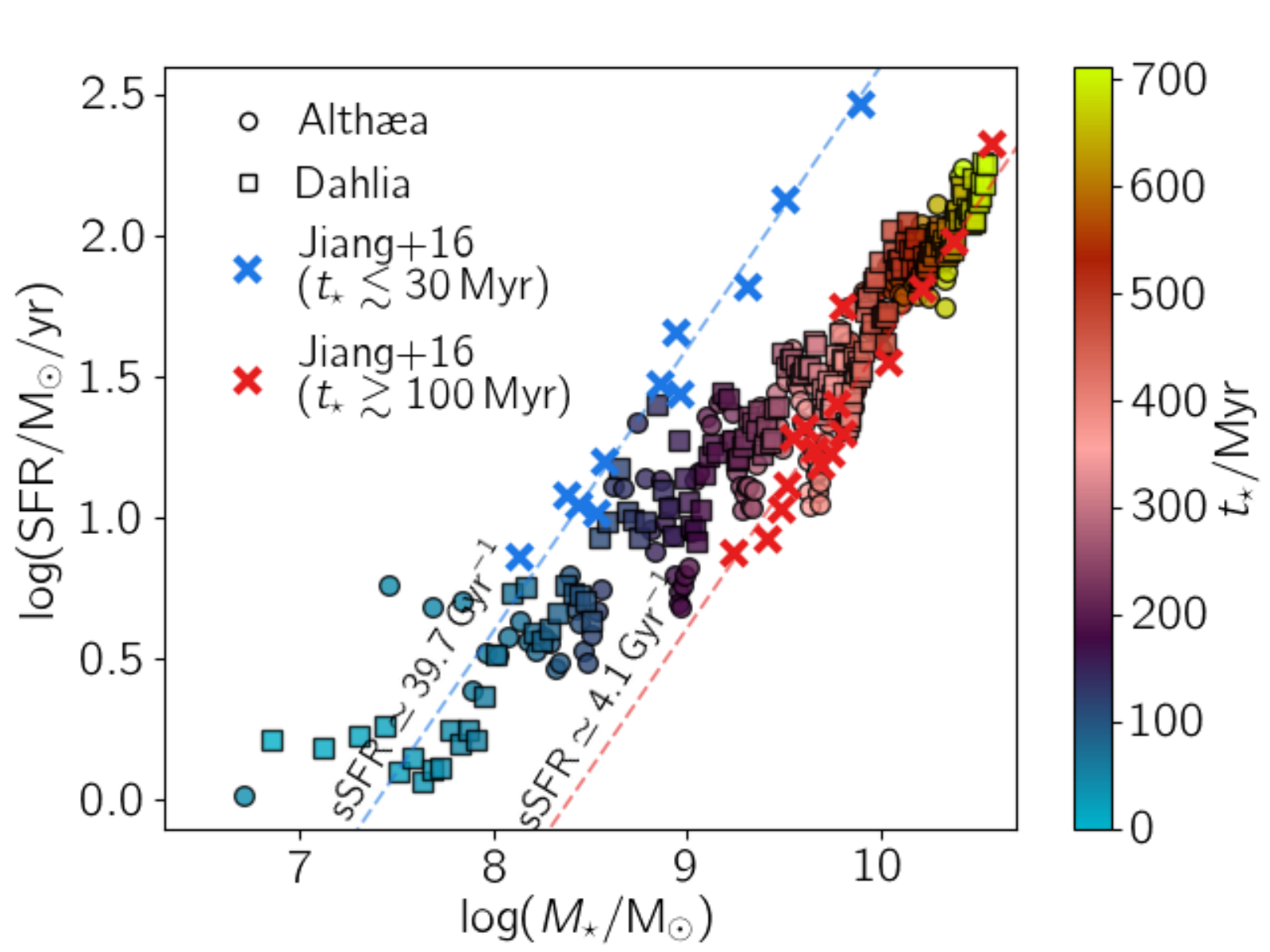}
\caption{
SFR vs stellar mass ($M_{\star}$) for Alth{\ae}a (circles) and Dahlia (squares), with symbols coloured accordingly to the age $t_{\star}$. With crosses we overplot SFR and $M_{\star}$ inferred from 27 galaxies observed at $z\simeq 6$ by \J16cit. Following \J16cit analysis\textsuperscript{\ref{foot_j16}}, galaxies identified as young and old are plotted in blue and red, respectively. To guide the eye, the linear correlation between the data sets are also shown with a dashed lines. See the text for more details.
\label{fig_sfr_mass_obs_comparison}
}
\end{figure}

In Fig.~\ref{fig_sfr_comparison} we plot the SFR history as a function of ``galaxy age'' ($t_{\star}$) for Dahlia and Alth{\ae}a; $t_{\star}=0$ marks the first star formation event in Dahlia\footnote{Note that even with the same modelling and IC, differences in the SFR may arise as a result of stochasticity in the star formation prescription eq. \ref{eq_sk_relation}. Such differences vanish once the SFR is averaged on timescales longer than the typical free-fall time of the star forming gas.}.
For both Dahlia and Alth{\ae}a the SFR has an increasing trend which can be approximated with good accuracy (within a factor of 2) as ${\rm SFR} = 1.5 \log(t_{\star}/(30\,\myr))$. However, on average the SFR in Dahlia is larger by a factor $\simeq 1.5 \pm 0.6$ when averaged over the entire SFR history ($\simeq 700\,\myr$). 
Thus, in spite of very different chemical prescriptions, the SFR in the two galaxies shows very little variation. Stated differently, the higher critical density for star formation arising from non-equilibrium chemistry does not alter significantly the rate at which stars form, as already noticed in Sec.~\ref{sec_chem_test}. This also entails a comparable metallicity, and we note that in both galaxy most of the metal mass is locked in stars (see Tab. \ref{tab_summary}), as they are typically formed from the most enriched regions.

It is interesting to check the evolutionary paths of Dahlia and Alth{\ae}a (Fig.~\ref{fig_sfr_mass_obs_comparison}) in the standard SFR vs. stellar mass ($M_{\star}$) diagram, and compare them with data\footnote{SFR and $M_{\star}$ have been derived by assuming an exponentially \textit{increasing} SFR, consistent with the history of both our simulated galaxies (Fig.~\ref{fig_sfr_comparison}).\label{foot_j16}} inferred from $z\simeq 6$ observations of 27 Lyman Alpha Emitters (LAE) and LBGs \citep[][hereafter \J16cit]{jiang:2016apj}. By using multi-band data, precise redshift determinations, and an estimate of nebular emission from Ly$\alpha$, \J16cit~were able to distinguish between a young ($t_{\star}\lsim 30\,\myr$) and an old ($t_{\star}\gsim 100\,\myr$) subsample. Each subsample exhibits a linear correlation in $\log {\rm SFR} - \log M_*$, albeit with a different normalization: the young (old) subsample has a ${\rm sSFR} = {\rm SFR}/M_{\star} = 39.7\, {\rm Gyr}^{-1} (4.1\, {\rm Gyr}^{-1})$. 

The SFR vs stellar mass of our simulated galaxies for $M_{\star}\lsim 10^{8.5} \msun$ ($t_{\star}\lsim 100\, \myr$) is fairly consistent with the young subsample relation (keeping in mind stochasticity effects at low stellar masses). At later evolutionary stages ($t_{\star}\gsim 300\, \myr$ or $M_{\star}\gsim 10^{9.5} \msun$), Dahlia and Alth{\ae}a nicely shift to the lower sSFR values characterizing the old \J16cit subsample data. This shift must be understood as a result of increasing stellar feedback: as galaxies grow, the larger energy input from the accumulated stellar populations hinders subsequent SFR events. 
Note that at late times ($t_{\star}\gsim 300$ Myr), when $M_{\star} = 5\times 10^9 M_\odot$, the sSFRs of Dahlia and Alth{\ae}a are in agreement with analytical results by \citet{behroozi:2013apj}, and with $z = 7$ observations by \citet{gonzalez:2010}.

\begin{figure}
\centering
\includegraphics[width=0.49\textwidth]{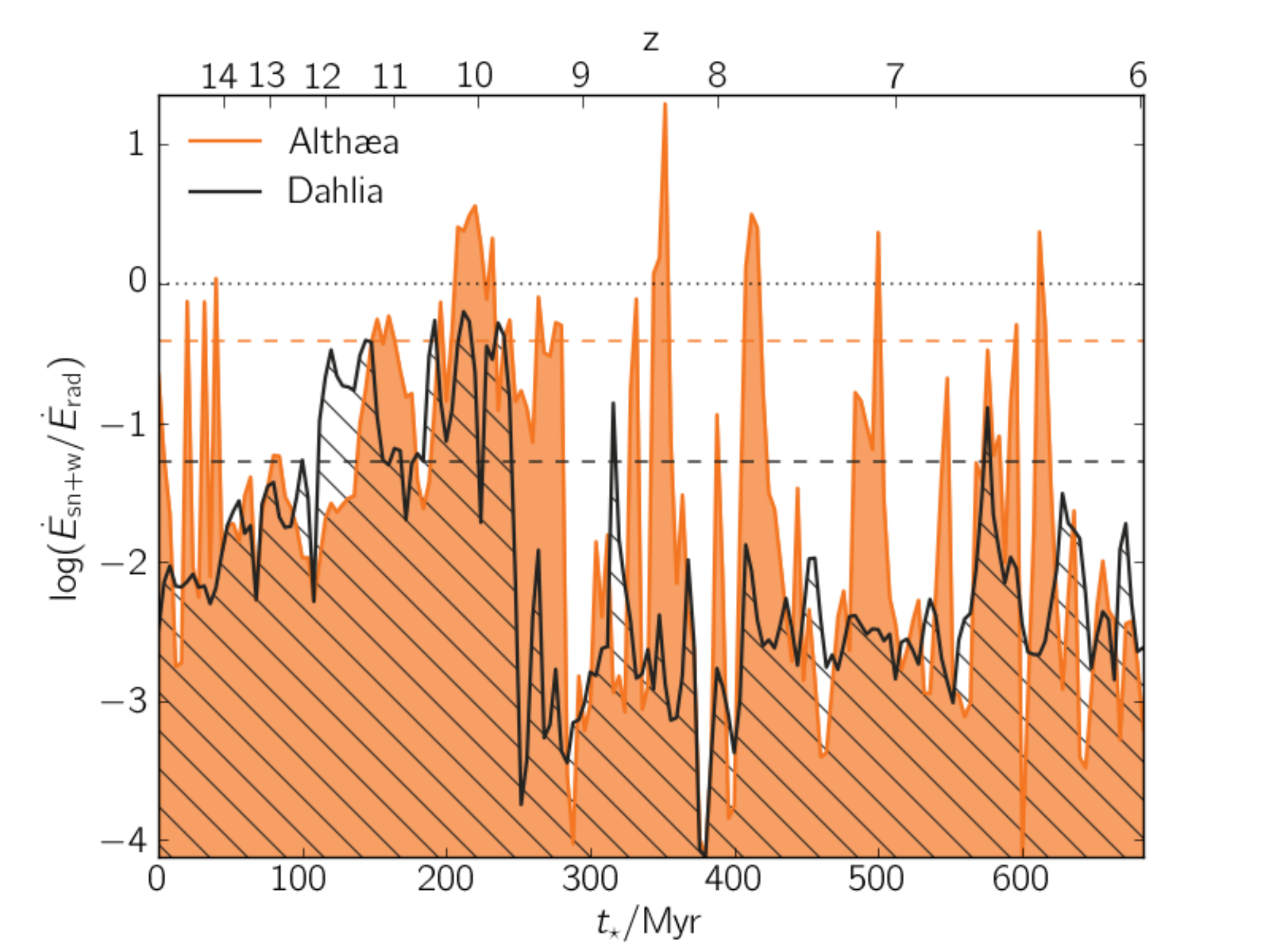}
\caption{
Ratio of mechanical ($\dot{E}_{\rm sn+w}$) and radiative ($\dot{E}_{\rm rad}$) energy deposition rates by stars as a function of galaxy age ($t_{\star}$) for Dahlia (black line/hatched area) and Alth{\ae}a (orange/transparent). Dashed lines indicate the $\simeq 700\,\myr$ time-averaged mean of the ratios for each galaxy. To guide the eye we plot the unity value (dotted grey line). Similar to Fig.~\ref{fig_sfr_comparison} the ratios are averaged over $4\,\myr$. The upper horizontal axis indicates redshift.
\label{fig_energy_comparison}
}
\end{figure}

\begin{table*}
\centering
\begin{tabular}{l|c|c|c|c}
\hline
Property                             & Symbol                              & Dahlia    & Alth{\ae}a & [units]                   \\
\hline
\hline
Star formation rate                  & $\rm SFR$                            & $156.19$ & $136.50$   & $\msun/{\rm yr}$          \\
Specific SFR                         & $\rm sSFR$                           & $4.45  $ & $5.23  $   & ${\rm Gyr}^{-1}$          \\
Stellar mass                         & $M_{\star}$                          & $3.51  $ & $2.61  $   & $10^{10}\msun$            \\
Metal mass in stars                  & $M_{\star}^Z$                        & $8.20  $ & $5.87  $   & $10^{8}\msun$             \\
\hline
Gas mass                             & $M_{g}$                              & $1.23 $  & $2.72  $   & $10^9\msun$               \\ 
H$_2$ mass                           & $M_{\rm H2}$                         & $17.01$  & $4.76  $   & $10^7\msun$               \\
Metal mass                           & $M_{Z}$                              & $1.41 $  & $2.48  $   & $10^7\msun$               \\
Disk radius                          & $r_d$                                & $610  $  & $504   $   & $\rm pc$                  \\
Disk scale height                    & $H$                                  & $224  $  & $191   $   & $\rm pc$                  \\
\hline
Gas density                          & $\langle n\rangle$                   & $23.89$  & $164.41$   & $\cc$                     \\
H$_2$ density                        & $\langle \nh2\rangle$                & $6.62 $  & $4.95$     & $\cc$                     \\
Metallicity                          & $\langle Z\rangle$                   & $0.57 $  & $0.46$     & $\zsun$                   \\
\hline
Gas surface density                  & $\langle\Sigma\rangle$               & $37.89$  & $222.02$   & $\msun/{\rm pc}^{2}$      \\
Star formation surface density       & $\langle\dot{\Sigma}_{\star}\rangle$ & $0.40 $  & $0.83  $   & $\msun/{\rm pc}^{2}/\myr$ \\
Luminosity \CII~$157.74\,\mu{\rm m}$ & $L_{\rm CII}$                        & $3.39 $  & $21.08 $   & $10^7\lsun$               \\
Luminosity \HH~$17.03\,\mu{\rm m}$   & $L_{\rm H2}$                         & $2.31 $  & $33.24 $   & $10^5\lsun$               \\
\end{tabular}
\caption{
Physical properties of Dahlia and Alth{\ae}a at $z=6$. The values refer to gas and stars within $2.5\,{\rm kpc}$ from the galaxy center (similar to the field of view in Fig.s \ref{fig_mappe_comparison_1} and \ref{fig_mappe_comparison_2}). The effective radius, $r_d$, and gas scale height, $H$, are calculated from the principal component analysis of the density field. Values for $n$, $\nh2$, $Z$, $\Sigma$, and $\dot{\Sigma}_{\star}$ represent mass-weighted averages.
\label{tab_summary}
}
\end{table*}

As feedback clearly plays a major role in the overall evolution of early galaxies, we turn to a more in-depth analysis of its energetics. 
This can be quantified in terms of the stellar energy deposition rates in \emph{mechanical} (SN explosions + OB/AGB winds\footnote{On average OB/AGB winds account only for $\lsim 10\%$ of the SN power.}, $\dot{E}_{\rm sn+w}$) and \emph{radiative} ($\dot{E}_{\rm rad}$) forms. These are shown as a function of time in Fig.~\ref{fig_energy_comparison}. The $\dot{E}_{\rm sn+w}/\dot{E}_{\rm rad}$ ratio shows short-term ($\simeq 20 \,\myr$) fluctuations corresponding to coherent burst of star formation/SN activity.

Barring this time modulation, on average the mechanical/radiative energy ratio increases up to $\simeq 250\,\myr$, when it suddenly drops and reaches an equilibrium value. This implies that radiation pressure dominates the energy input; consequently it represents the major factor in quenching star formation. While this is true throughout the evolution, it becomes even more evident after $\simeq 250\,\myr$, when the first stellar populations with $Z_{\star}\gsim 10^{-1}\zsun$ enter the AGB phase. At that time, winds from AGBs enrich their surroundings with metals and dust. As dust produced by AGBs remains more confined than SN dust around the production region, it provides a higher opacity, thus boosting radiation pressure via a more efficient dust-gas coupling (see also \P17cit).

For Dahlia the radiative energy input rate is about 20 times larger than the mechanical one, while for Alth{\ae}a such ratio is on average $8$ times higher, although larger fluctuations are present. The latter are caused by the occurrence of more frequent and powerful bursts of SN events in Alth{\ae}a. Why does this happen? 

The answer has to do with the different gas morphology. As already noted discussing Fig.~\ref{fig_chimica_krome}, the higher critical density for star formation imposed by non-equilibrium chemistry has a number of consequences: (a) each formation event can produce a star cluster with an higher mass;
(b) star formation is more likely hosted in isolated high density clumps (see later, particularly Fig.~\ref{fig_morfologia}); (c) in a clumpier disk, SN explosions can easily break into more diffuse regions. The combination of (a) and (b) increases the probability of spatially coherent explosions having a stronger impact on the surrounding gas; due to (c), the blastwaves suffer highly reduced radiative losses \citep{gatto:2015mnras}, and affect larger volumes. 
Similar effects have been also noted in the context of single giant MCs ($\sim10^6\msun$), where unless the SNs explode coherently, their energy is quickly radiated away because of the very high gas densities \citep{reyraposo:2017}.
For the reminder of the work we focus on $z=6$, when the galaxies have an age of $t_{\star}\simeq 700\, \myr$.

\subsection{Galaxy morphology}\label{sec_morphology}

\begin{figure*}
~\hfill\includegraphics[width=0.96\textwidth]{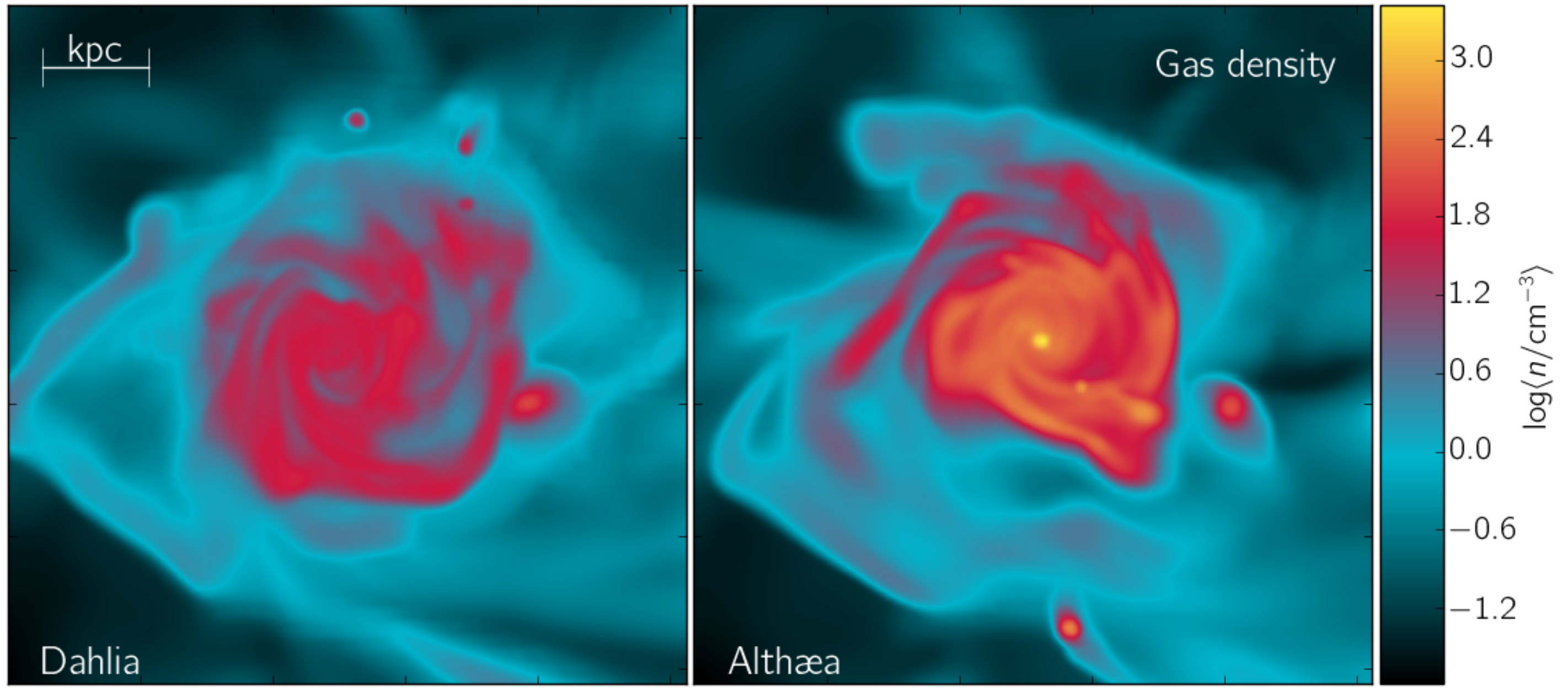}\hfill~\\	
~\hfill\includegraphics[width=0.96\textwidth]{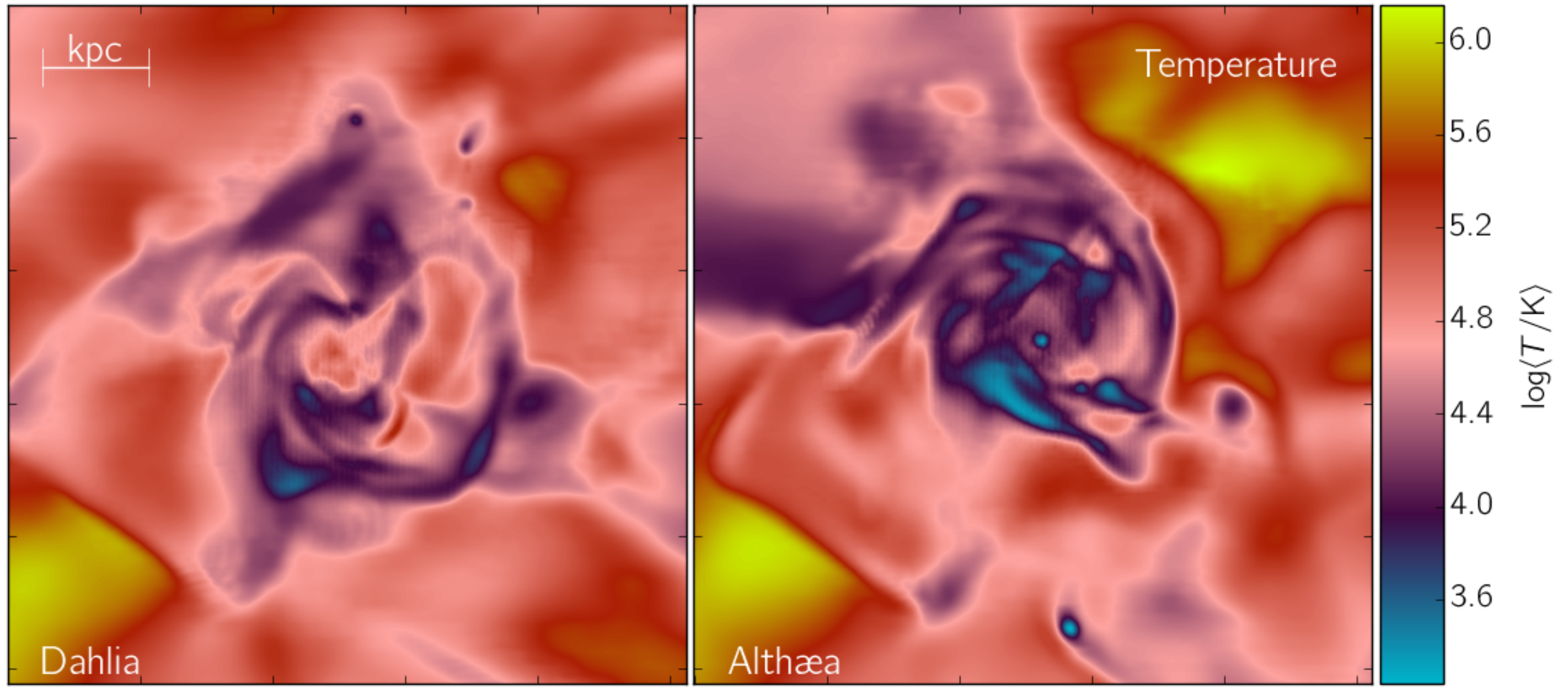}\hfill~\\
~\hfill\includegraphics[width=0.96\textwidth]{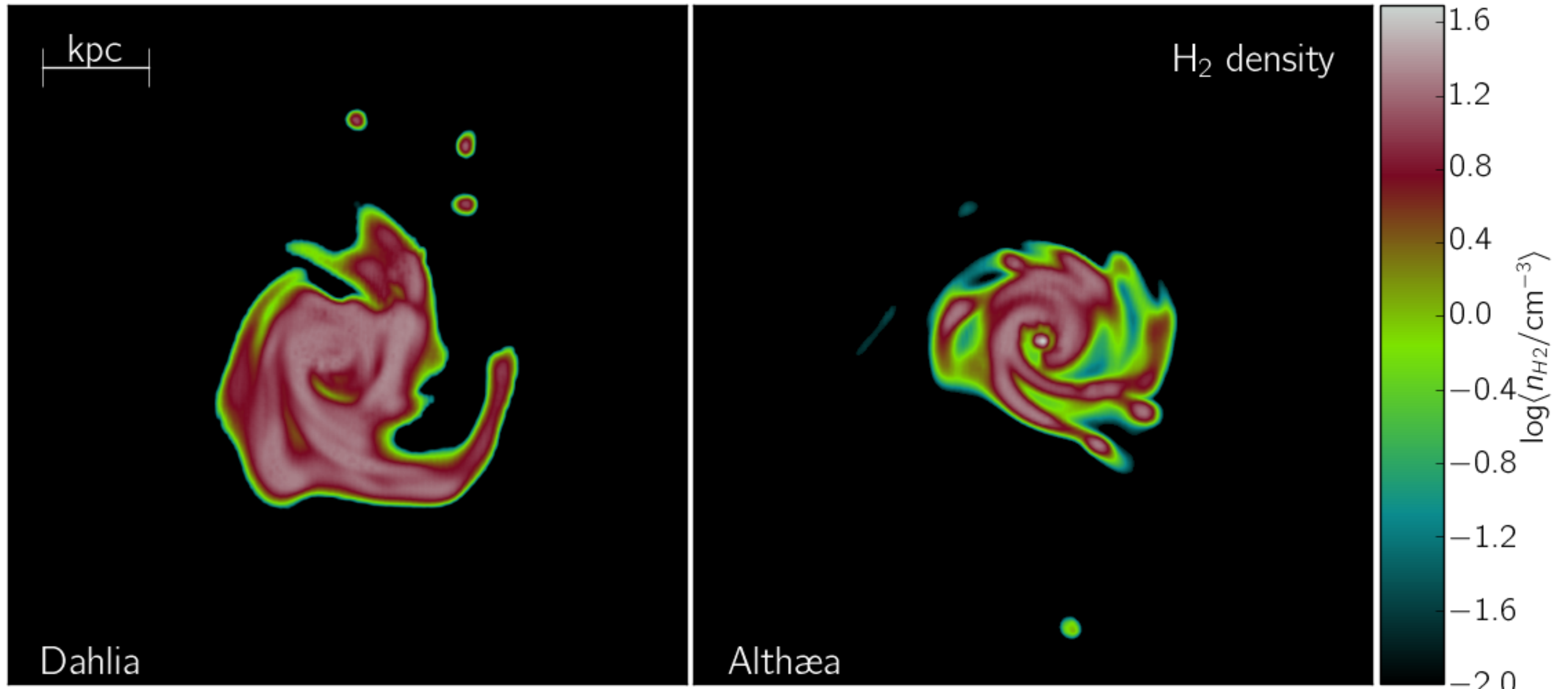}\hfill~\\
\caption{
(Caption next page.) %
\label{fig_mappe_comparison_1}
}
\end{figure*}

\addtocounter{figure}{-1}
\begin{figure*}
\caption{(Previous page.)
Face-on maps\textsuperscript{\ref{footnote_pymses}} of Dahlia (left panels) and Alth{\ae}a (right) at age $t_{\star}\simeq 700\, \myr$ ($z=6$). Shown are line-of-sigh mass weighted average of the gas density (upper panels), temperature (middle), and \HH~density (lower) fields with amplitude given by the colorbar. The maps are $6.31\,{\rm kpc}$ on a side. 
}
\end{figure*}

\begin{figure}
\centering
\includegraphics[width=0.49\textwidth]{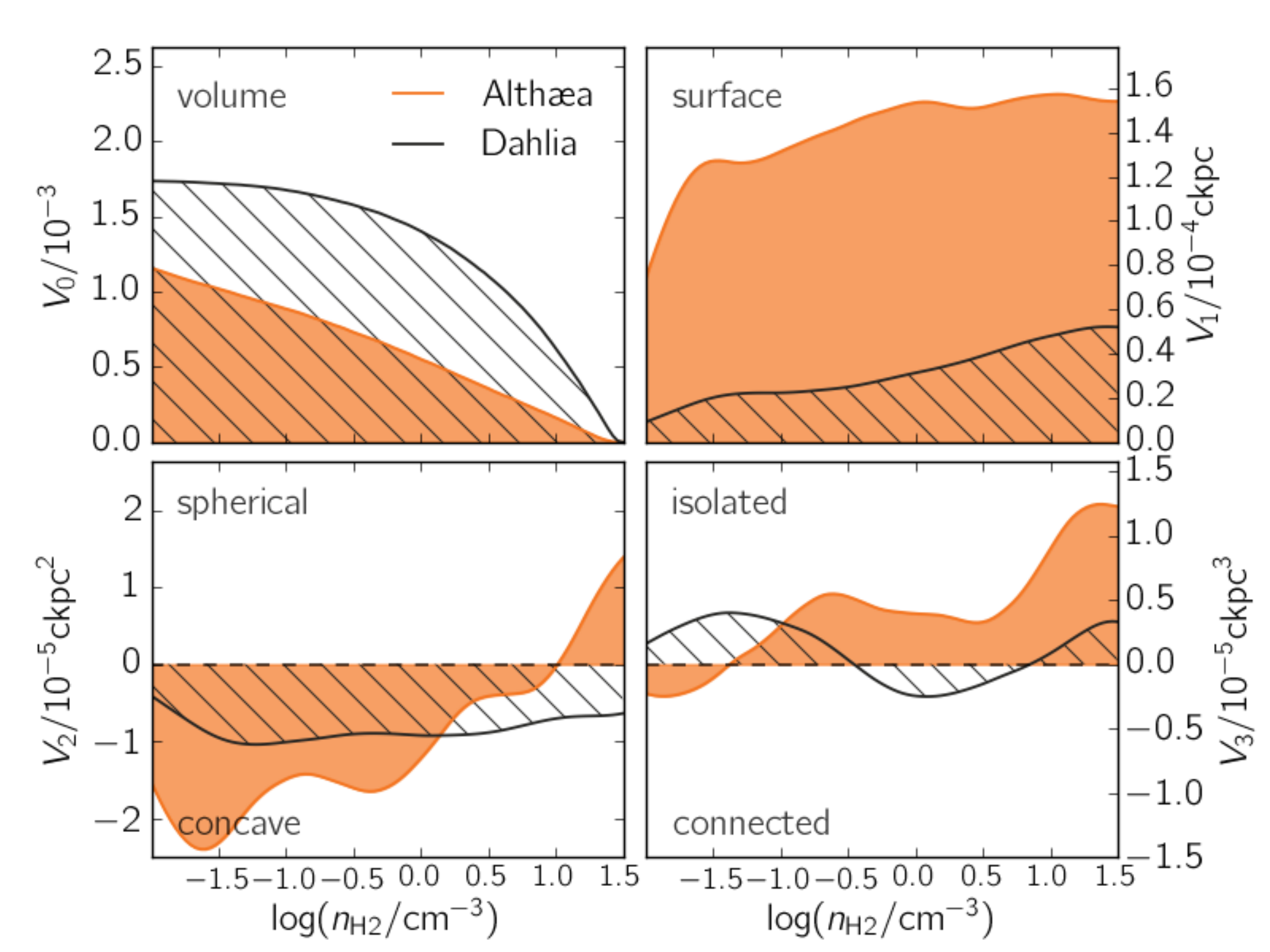}
\caption{
Morphological comparison of the molecular gas at $z=6$.
In the four panels we plot the Minkowsky functionals ($V_{0},V_{1},V_{2},V_{3}$) of the \HH~density field ($\nh2/\cc$).
Functionals are plotted with black line and hatched regions for Dahlia, with orange line and transparent region for Alth{\ae}a.
Note that Minkowsky functionals are indicated in comoving units.
For detail on the calculation of the Minkowsky functional see App.~\ref{sec_app_minchioschi} (in particular see Fig.~\ref{fig_morfologia_test}).
\label{fig_morfologia}
}
\end{figure}

Dahlia and Alth{\ae}a sit at the centre of a cosmic web knot and accrete mass from the intergalactic medium mainly via 3 filaments of length $\simeq 100\,{\rm kpc}$. In both simulations, the large scale structure is similar, and we refer the reader to Sec.~3.1 of \P17cit for its analysis. Differences between the simulation are expected to arise on the ISM scale, whose structure is visible on $\simeq 7\,{\rm kpc}$ scales. In Fig.~\ref{fig_mappe_comparison_1} we show the gas density, temperature, and \HH~density ($\nh2 = \fh2\,n\,\mu$) fields for Dahlia and Alth{\ae}a. The map\footnote{The maps of this work are rendered by using a customized version of \textlcsc{pymses} \citep{labadens:2012aspc}, a visualization software that implements optimized techniques for the AMR of \textlcsc{ramses}.\label{footnote_pymses}} centers coincide with Dahlia's stellar center-of-mass. 

\subsubsection{Overview}

Qualitatively, both galaxies show a clearly defined, albeit somewhat perturbed, spiral disk of radius $\simeq 0.5\,\rm kpc$, embedded in a lower density ($n\simeq 0.1\,\cc$) medium. However the mean disk gas density for Dahlia is $\langle n\rangle=24\,\cc$, while for Alth{\ae}a $\langle n\rangle = 164\,\cc$ (see Tab.~\ref{tab_summary}). The temperature structure shows fewer differences, i.e. the inner disk is slightly hotter for Dahlia ($T\simeq 300\,\rm K$) than for Alth{\ae}a ($T\simeq 100\,\rm K$), which features instead slightly more abundant and extended pockets of shock-heated gas ($T\gsim 10^6$). Such high-$T$ regions are produced by both accretion shocks and SN explosions. In both cases the typical \HH~density is the same, i.e. $\langle \nh2\rangle = 5\,\cc$, however, with respect to Dahlia, Alth{\ae}a shows a slightly smaller disk, that also seems more clumpy. 

To summarize, the galaxies differ by an order of magnitude in atomic density, but have the same molecular density. In spite of this difference, the SFR are roughly similar. This can be explained as follows. To first order, in our model $SFR \propto \nh2 n^{1/2} V$, where $V = 2\pi r_d^2 H$ is the galaxy volume (Tab.~\ref{tab_summary}). It follows that the larger density is largely compensated by the smaller Alth{\ae}a volume.

\subsubsection{In-depth analysis}

Fig.~\ref{fig_mappe_comparison_1} visually illustrates the morphological differences between the two galaxies. The gas in Alth{\ae}a appears clumpier than in Dahlia. To quantify this statement we start by introducing the \HH~clumping factor on the smoothing scale $r$, which is defined as\footnote{To calculate the clumping factor, first we construct the 3D unigrid cube of the \HH~mass field, then we smooth it with a Gaussian kernel of scale $r$ and finally we calculate the mass-weighted average and variance of the smoothed \HH~density field.}
\be
C(r) = \langle \nh2^{2} \rangle_{r}/\langle \nh2 \rangle_{r}^{2}\,,
\ee
For Dahlia $C(r)$ decreases from $10^3$ to 10 going from 30 pc to 1 kpc, while for Alth{\ae}a $C(r)$ is $\gsim 2$ times larger 
on all scales.

A more in-depth analysis can be performed using the Minkowsky functionals \citep[][App.~\ref{sec_app_minchioschi}]{schmalzing:1998,gleser:2006mnras,yoshiura:2017mnras} which can give a complete description of the molecular gas morphological structure. For a 3-dimensional field, 4 independent Minkowsky functionals can be defined. Each of the functionals, $V_{i}(\nh2) \, (i=0,\dots,3)$ characterizes a different morphological property of the excursion set with \HH~density $>\nh2$: $V_{0}$ gives the volume filling factor, $V_{1}$ measures the total enclosed surface, $V_{2}$ is the mean curvature, quantifying the sphericity/concavity of the set, and $V_{3}$ estimates the Euler characteristic (i.e. multiple isolated components vs. a single connected one. Appendix \ref{sec_app_minchioschi} gives more rigorous definitions with an illustrative application (Fig.~\ref{fig_morfologia_test}).

In Fig.~\ref{fig_morfologia} we plot the Minkowsky functionals ($V_{0},V_{1},V_{2},V_{3}$) calculated for the \HH~density field for Dahlia and Alth{\ae}a.
The $V_{0}$ functional analysis shows that Alth{\ae}a is more compact, i.e. for each $\nh2$ value Dahlia's excursion set volume is larger and it plummets rapidly at large densities. On the other hand, the set surface of Alth{\ae}a is larger by about a factor of 5, implying that this galaxy is fragmented into multiple, disconnected components. 
This is confirmed also by Alth{\ae}a's larger ($10\times$) Euler characteristic measure, $V_{3}$, an indication of the prevalence of isolated structures. This feature becomes more evident towards larger densities, as expected if \HH~is concentrated in molecular clouds\footnote{$V_{3}>0$ values at $\log(\nh2/\cc)\simeq 1.2$ in Dahlia are mainly due to the presence of the 3 satellites/clumps outside the disk.}.

Further, in Dahlia most of the molecular gas resides in connected ($V_{3}\lsim 0$) disk regions, with a concave shape ($V_{2}<0$). For Alth{\ae}a there is a transition: for $\log(\nh2/\cc)\lsim 1$ the gas has a concave ($V_{2}<0$), disjointed ($V_{3}>0$), filamentary structure, while for $\log(\nh2/\cc)\gsim 1$ the galaxy is composed by spherical clumps ($V_{2}>0$).

\subsection{ISM thermodynamics}\label{sec_thermo_state}

\begin{figure*}
\centering
\includegraphics[width=0.49\textwidth]{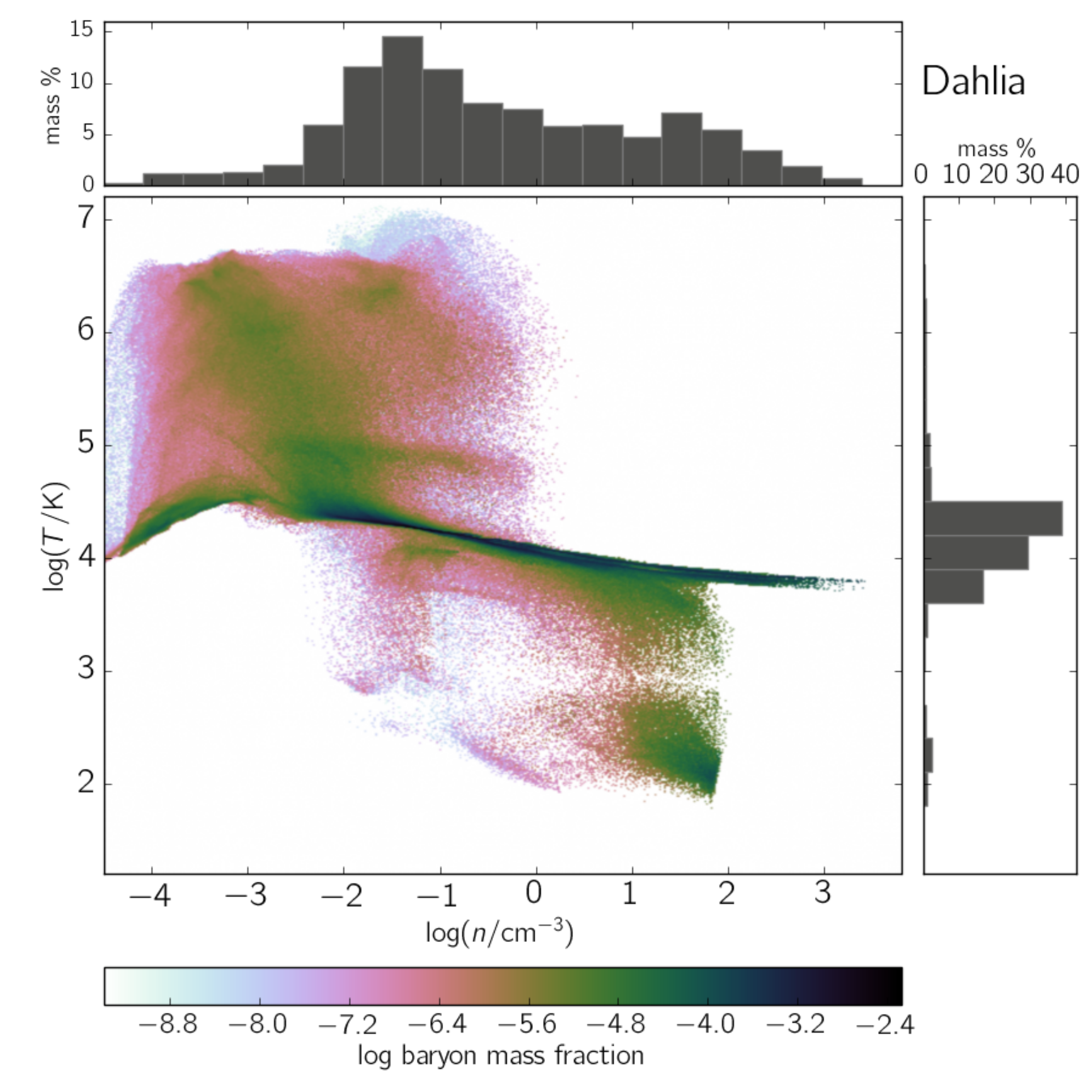}
\includegraphics[width=0.49\textwidth]{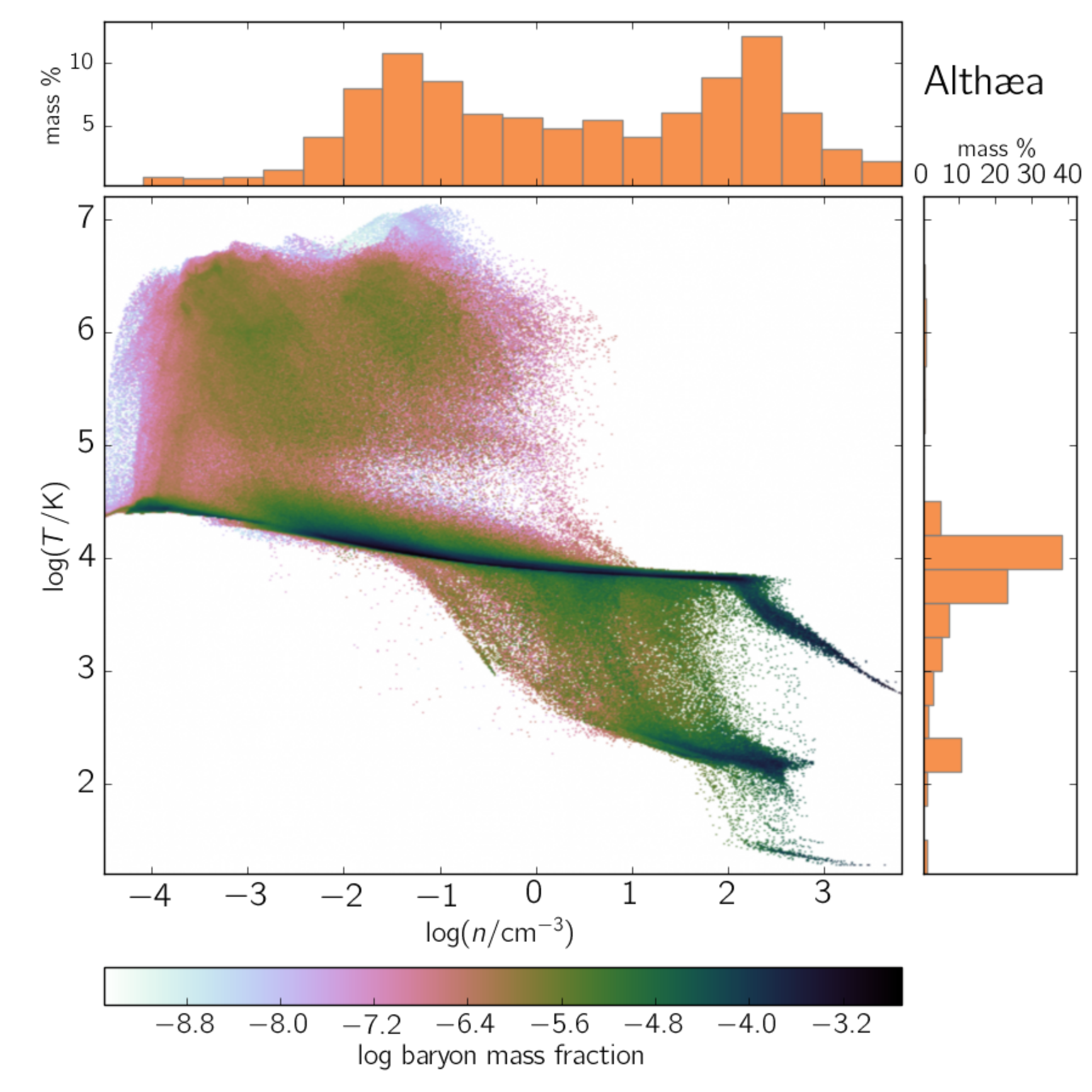}
\caption{
Equation of state (EOS) of the gas within $30\,{\rm kpc}$ for Dahlia (left panel) and Alth{\ae}a (right panel) at $t_{\star}\simeq 700\, \myr$ ($z=6$).
EOS are shown as mass-weighted probability distribution function (PDF) in the density-temperature ($n-T$) plane, as specified by the colorbar. For both galaxies, the EOS projection on the $n$ ($T$) axis is additionally shown as an horizontal (vertical) inset. The 2D EOS are normalized such that the integral on the $n-T$ plane is unity; the projected EOS are normalized such that the sum of the bins is equal to $100\%$.
\label{fig_eos}
}
\end{figure*}

\begin{figure*}
\centering
\includegraphics[width=0.49\textwidth]{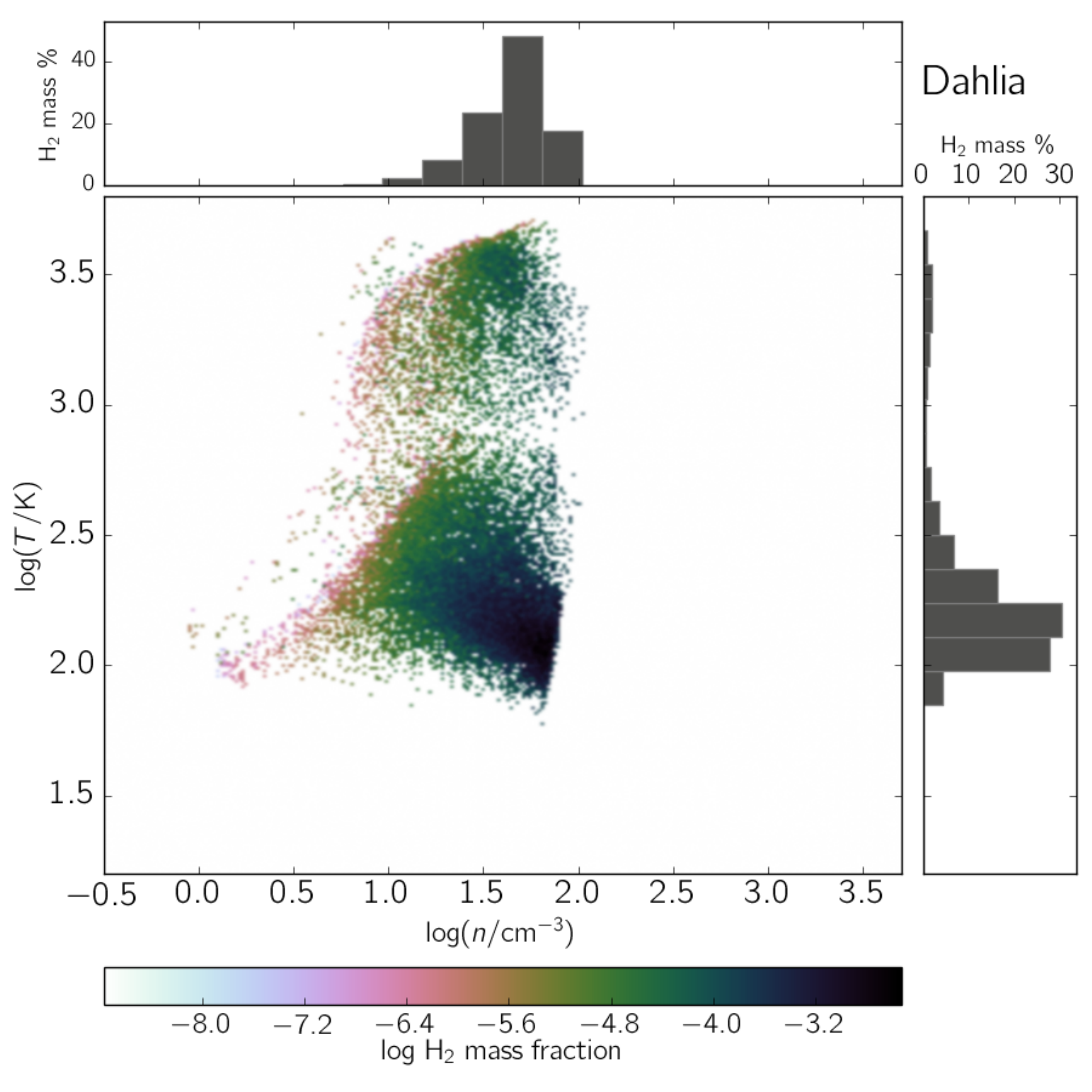}
\includegraphics[width=0.49\textwidth]{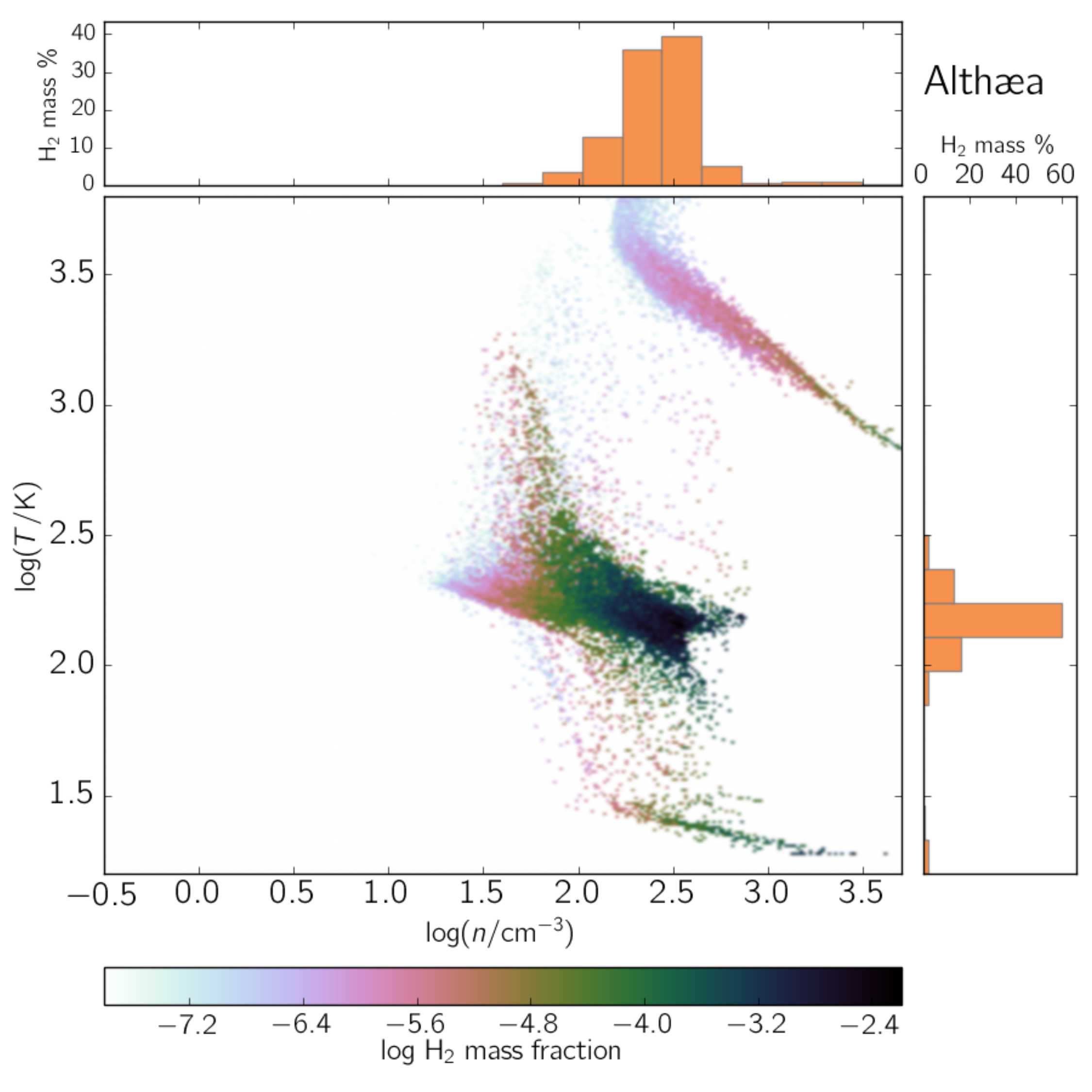}
\caption{
EOS of the molecular (\HH) gas for Dahlia (left panel) and Alth{\ae}a (right panel) i.e. the \HH~mass-weighted PDF in the $n-T$ plane. Notation is similar to Fig.~\ref{fig_eos}, albeit a different region of $n-T$ plane is shown.
\label{fig_eos_h2}
}
\end{figure*}

The thermodynamical state of the ISM can be analyzed by studying the probability distribution function (PDF) of the gas in the density-temperature plane, i.e. the equation of state (EOS). In Fig.~\ref{fig_eos} we plot the mass-weighted EOS for Dahlia and Alth{\ae}a at $z=6$. We include gas within $30\,{\rm kpc}$, or $\simeq 2\, r_{\rm vir}$, from the galaxy center.

From the EOS we can see that in both galaxies $70\%$ of the gas in a photoionized state ($T\sim 10^4\rm K$), that in Dahlia is induced by the \citet[][]{Haardt:2012} UVB, while in Alth{\ae}a is mainly due to photo-electric heating on dust grains illuminated from the uniform ISRF of intensity $G$. Only $\simeq 10\%$ of the gas is in a hot $10^6$ K component produced by accretion shocks and SN explosions. A relatively minor difference descends from Alth{\ae}a's more effective mechanical feedback, already noted when discussing Fig.~\ref{fig_energy_comparison}: small pockets of freshly produced very hot ($\ge 10^6\,\rm K$) and diffuse ($0.1\, \cc$) gas are twice more abundant in Alth{\ae}a, as it can be appreciated from a visual inspection of the temperature maps in Fig.~\ref{fig_mappe_comparison_1}. 

Fig.~\ref{fig_eos} (in particular compare the upper horizontal panels) shows that the density PDF is remarkably different in the two galaxies. In Dahlia the distribution peaks at $0.1\,\cc$; Alth{\ae}a instead features a bi-modal PDF with a second, similar amplitude peak at $n\simeq 100\,\cc$. This entails the fact that the dense $\gsim 10\,\cc$ gas is about 2 times more abundant in the latter system. In addition, the very dense gas ($n\gsim 300\,\cc$), only present in Alth{\ae}a, can cool to temperatures of 30 K, not too far from the CMB one.

The high-density part of the PDF is worth some more insight as it describes the gas that ultimately regulates star formation. This gas is largely in molecular form, and accounts (see Tab.~\ref{tab_summary}) for $1.7\%$ ($13.8\%$) of the total gas mass in Dahlia (Alth{\ae}a). Its \HH~density-weighted distribution in the $n-T$ plane is reported in Fig.~\ref{fig_eos_h2}.
On average, the \HH~gas in Dahlia is 10 times less dense than in Alth{\ae}a as a result of the new non-equilibrium prescription requiring higher gas densities to reach the same $\fh2$ fraction; at the same time the warm ($T\gsim 10^{3}\rm K$) \HH~fraction drops from $20\%$ (Dahlia) to an almost negligible value. Clearly, the warm component was a spurious result as (a) \HH~cooling was not included, and (b) $\fh2$ was considered to be independent of gas temperature (see eq.s \ref{eq_fh2_anal}).
Note that in Alth{\ae}a traces of warm \HH~are only found at large densities, in virtually metal-free gas in which \HH~production must proceed via much less efficient gas-phase reactions rather than on dust surfaces. This tiny fraction of molecular gas can survive only if densities large enough to provide a sufficient \HH~self shielding against photodissociation are present.

Finally, the sharp EOS cutoff at $n\gsim 10^{2}\cc$ in Dahlia is caused by the density-threshold behavior mimicked by the enforced chemical equilibrium: above $n_{c} \simeq 26.45 \, (Z/\zsun)^{-0.87} \cc$ (Sec.~\ref{sec_chem_eq}) the gas is rapidly turned into stars. This spurious effect disappears in Alth{\ae}a, implementing a full time-dependent chemical network. 

\section{Observational properties}\label{sec_obs_prop}

As we already mentioned, the strongest impact of different chemistry implementations is on the gas properties, and consequently 
on ISM-related observables. In the following, we highlight the most important among these aspects.

\begin{figure}
\centering
\includegraphics[width=0.49\textwidth]{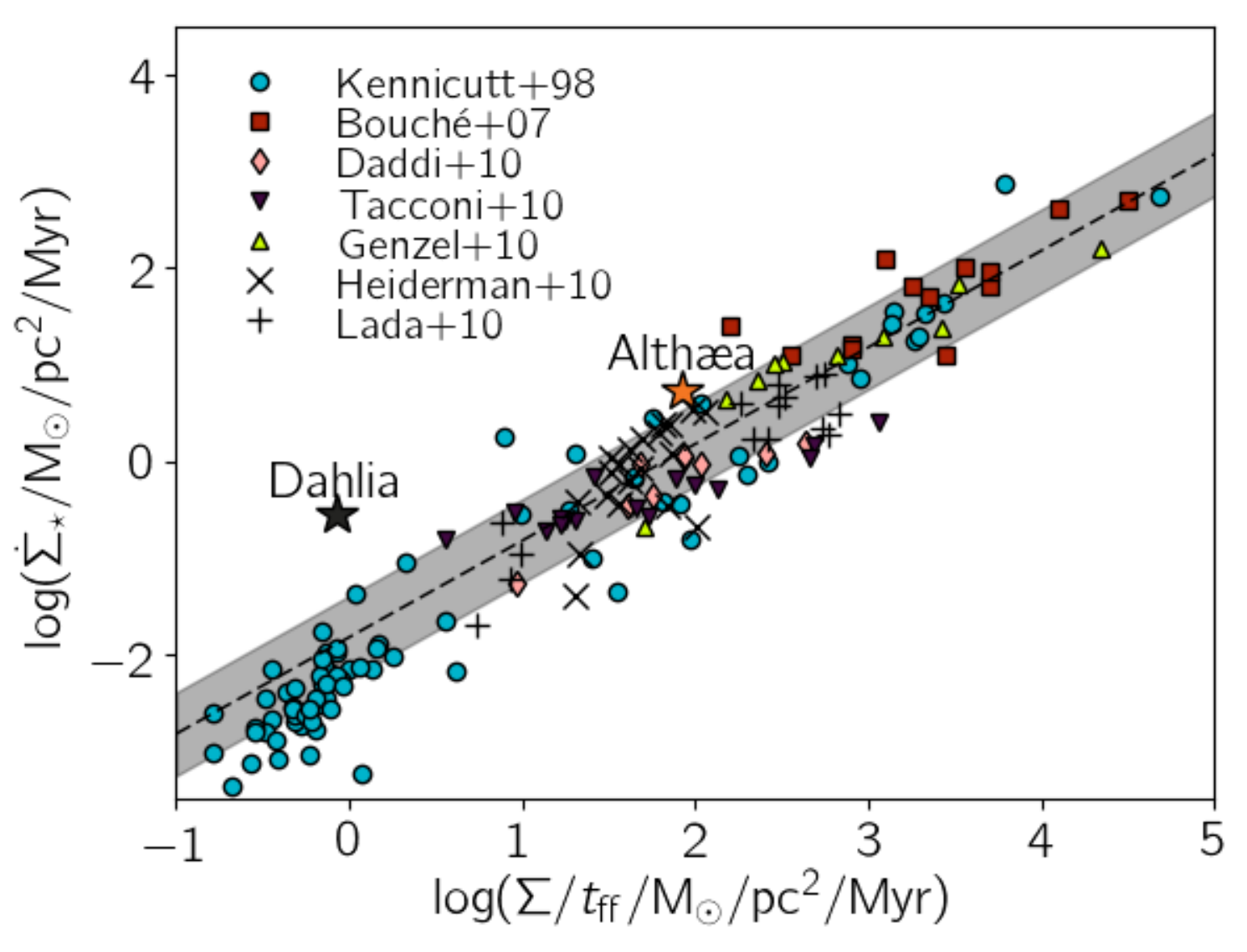}
\caption{
Comparison of the observed and simulated Schmidt-Kennicutt relation expressed in terms of $\dot{\Sigma}_{\star}$ - $\Sigma/t_{\rm ff}$.
Observations are taken from single MCs \citep{heiderman:2010,lada:2010apj}, local unresolved galaxies \citep{kennicutt:1998apj}, and moderate redshift unresolved galaxies \citep{bouche2007apj,daddi:2010apj,daddi:2010b,tacconi:2010Natur,genzel:2010mnras}; the correlation (dispersion) for the observation found by \citet[][see the text for details]{krumholz:2012apj} is plotted with a black dashed line (grey shaded region).
Dahlia and Alth{\ae}a averaged value are plotted with black and orange stars, respectively (see Fig.~\ref{fig_eos_sk} for the complete distribution in the simulated galaxies).
\label{fig_riassunto_sk}
}
\end{figure}

\begin{figure*}
\centering
\includegraphics[width=0.49\textwidth]{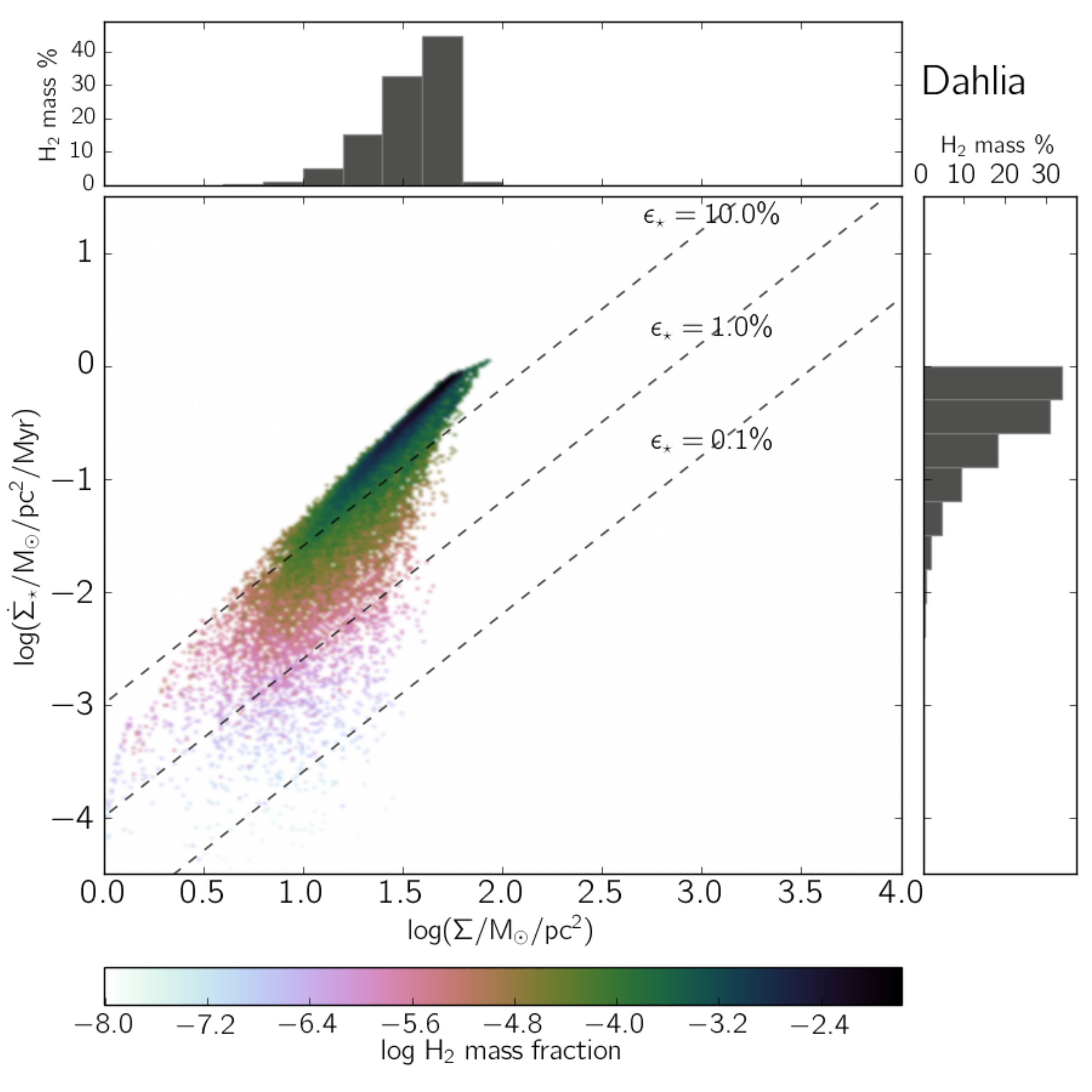}
\includegraphics[width=0.49\textwidth]{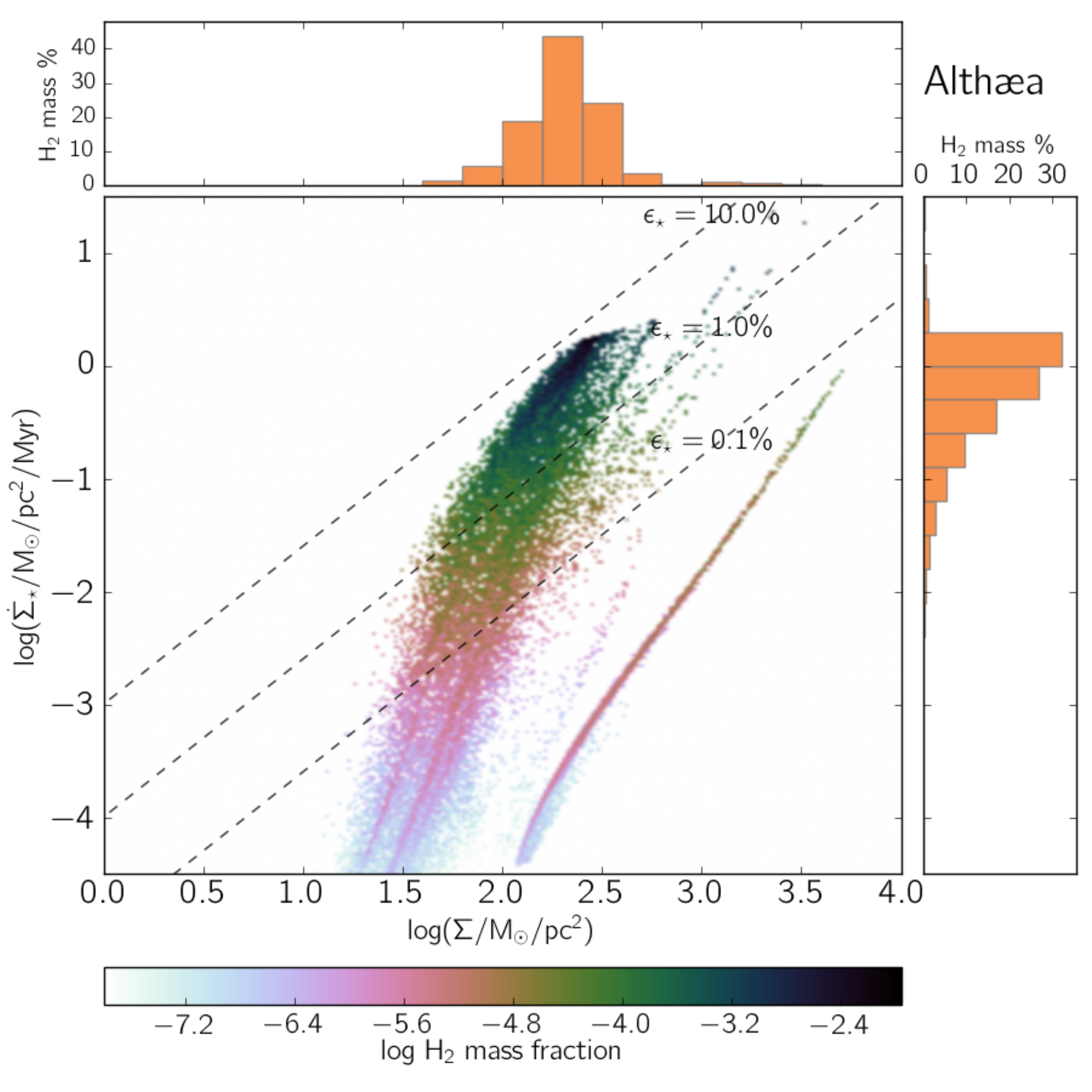}
\caption{
Schmidt-Kennicutt relation in Dahlia (left panel) and Alth{\ae}a (right panel) at $t_{\star}\simeq 700\, \myr$ ($z=6$).
The relation is plotted using the \HH~mass weighted PDF of the instantaneous SFR surface density $(\dot{\Sigma}_{\star}/\msun\,{\rm pc}^{-2}\,\myr^{-1})$ versus the total gas surface density ($\Sigma/\msun/{\rm pc}^{2}$).
On both panels with dashed grey lines we overplot the relation observed from \citet{Kennicutt:2012}, i.e. $\dot{\Sigma}_{\star} \propto \Sigma^{1.4}$, for several normalizations that written inline.
Otherwise notation is similar to Fig.s \ref{fig_eos_h2} and \ref{fig_eos}.
\label{fig_eos_sk}
}
\end{figure*}

\begin{figure*}
~\hfill\includegraphics[width=0.97\textwidth]{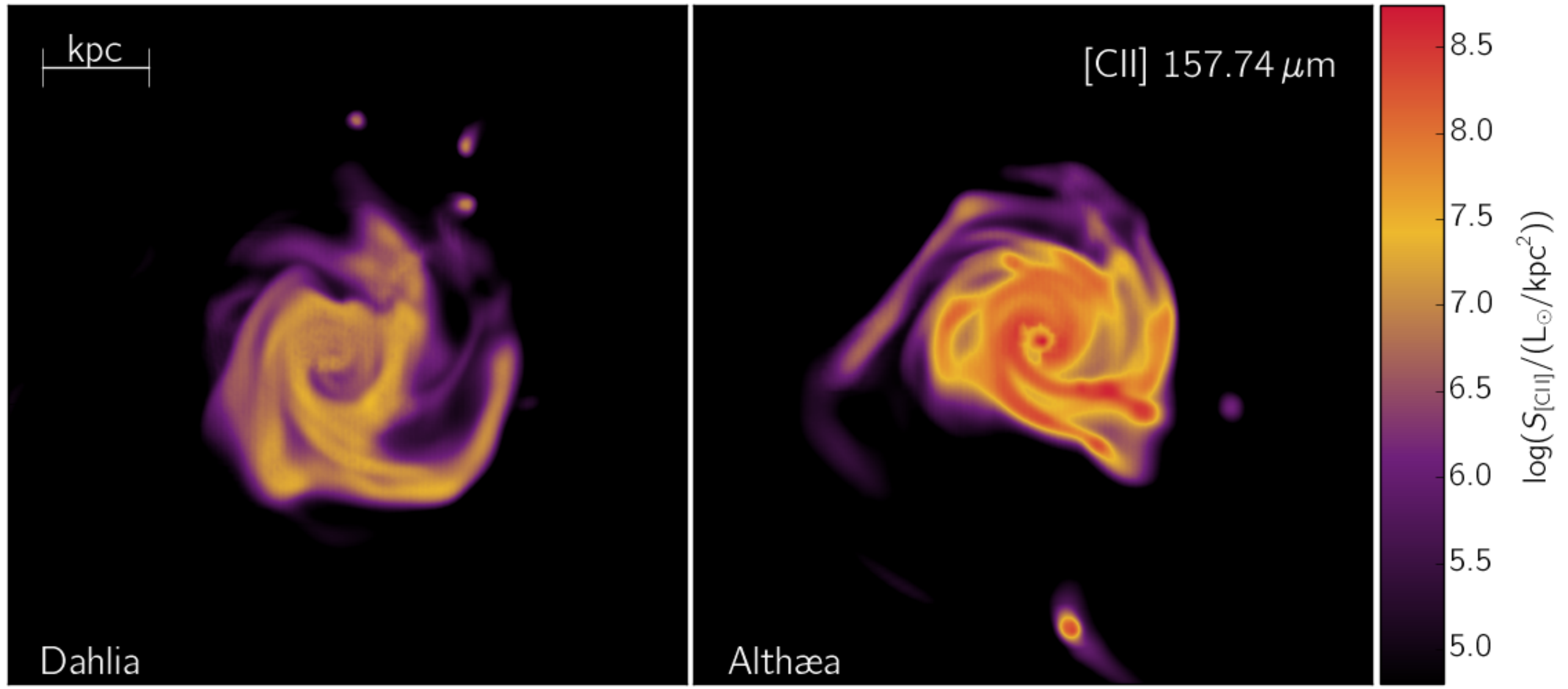}\hfill~\\
~\hfill\includegraphics[width=0.97\textwidth]{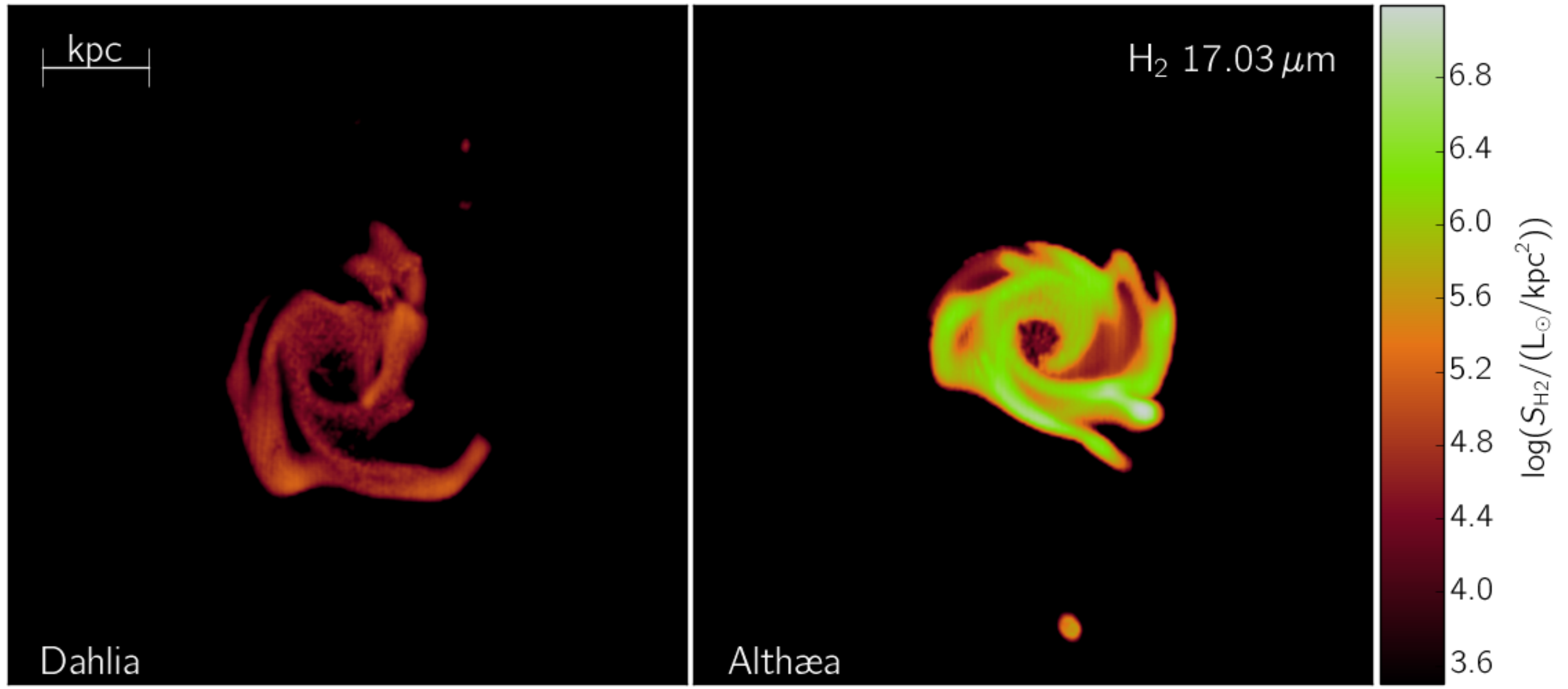}\hfill~\\
\caption{
Synthetic emission maps\textsuperscript{\ref{footnote_pymses}} of the simulated galaxies Dahlia (left panels) and Alth{\ae}a (right panels) at age $t_{\star}\simeq 700\, \myr$ ($z=6$).
Integrated surface brightness of \CII~($S_{\rm [CII]}/(\surfl)$) and \HH~($S_{\rm H2}/(\surfl)$) are shown in the upper and lower panels, respectively.
The field of view is the same as in Fig.~\ref{fig_mappe_comparison_1}.
\label{fig_mappe_comparison_2}
}
\end{figure*}

\begin{figure}
\includegraphics[width=0.49\textwidth]{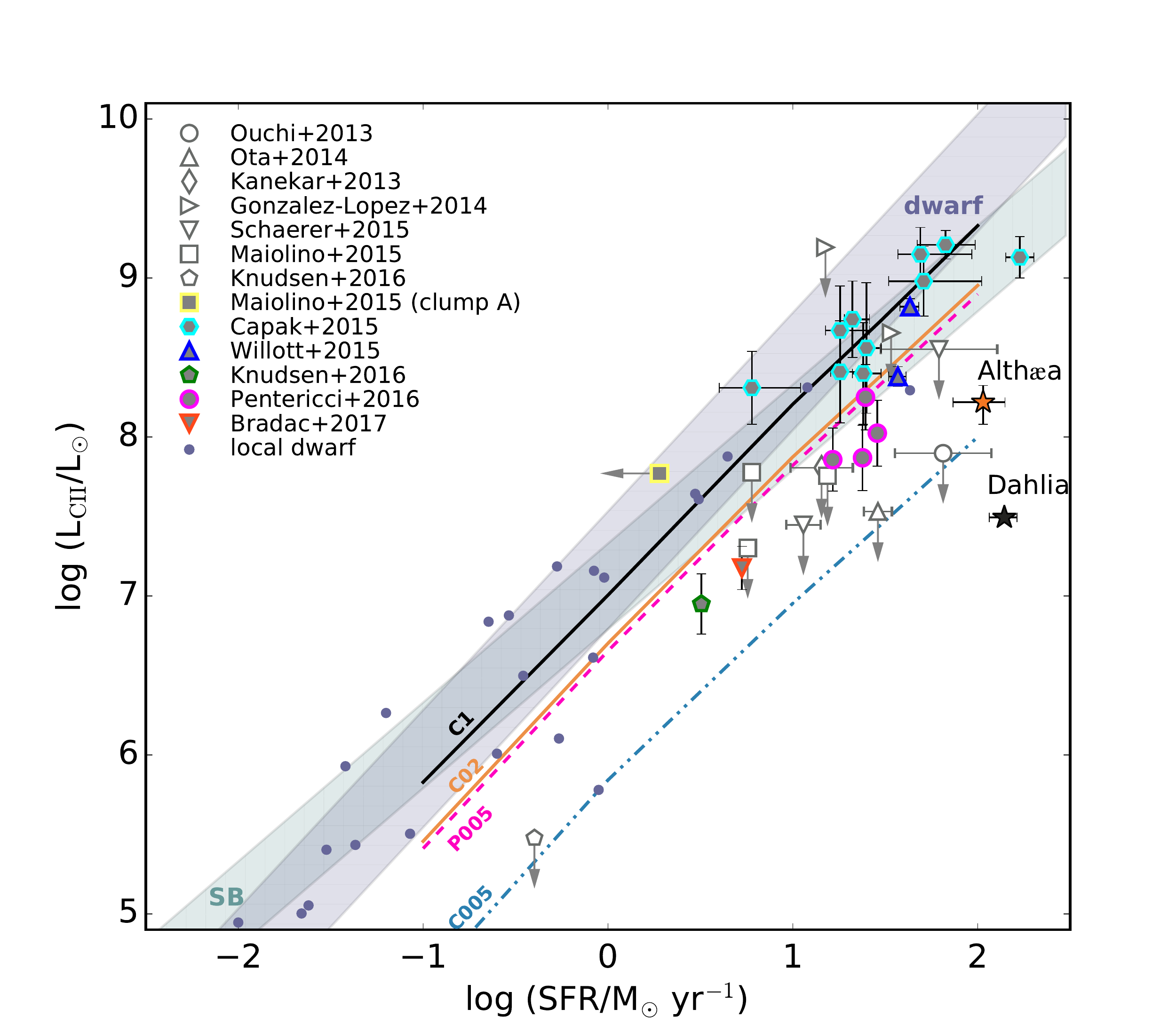}
\caption{
The \CII-SFR relation. Shown are Alth{\ae}a (orange star) and Dahlia (black) at 700 Myr or $z=6$; the errors refers to r.m.s. variation in the last $50\,\myr$.
Lines refer to results from the \V15cit~model: constant metallicity models with $Z=\zsun$ (solid black), $Z=0.2\,\zsun$ (solid orange), $Z=0.05\,\zsun$ (pink dashed), and a model with mean $\langle Z/\zsun \rangle = 0.05$ + density-metallicity relation extracted from cosmological simulations \citep[][blue dot-dashed]{pallottini:2014cgmh}.
Data for local dwarf galaxies \citep{delooze:2014aa} are plotted with little circles and the grey hatched region gives the mean and r.m.s. variation in the sample.
For high$-z$ galaxies, detections (upper-limits) are plotted with filled (empty) symbols, according to the inset legend. The high$-z$ sample include individual galaxies as BDF-3299 \citep{maiolino:2015arxiv,carniani:2017bdf3299}, HCM6A \citep{kanekar2013}, Himiko \citep{ouchi2013,ota:2014apj}, IOK-1 \citep{ota:2014apj}, and data from \citet[][$z\simeq 5.5$]{capak:2015arxiv}, \citet[][$z\simeq 6$]{willott:2015arxiv15}, \citet[][$z\simeq 7$]{schaerer:2015}, \citep[][$z\simeq 7$]{pentericci:2016apj}, \citet[][$z\simeq 8$]{gonzalezlopez:2014apj}, and lensed $z\simeq 6.5$ galaxies from \citet{knudsen:2016} and \citet[][]{bradac:2017}.
\label{fig_cii_sfr}
}
\end{figure}

\subsection{Schmidt-Kennicutt relation}
We start by analyzing the classical Schmidt-Kennicutt (SK) relation. This comparison should be interpreted as a consistency check of the balance between SF and feedback, since in the model we assume a SFR law that mimics a SK relation (eq. \ref{eq_sk_relation}).

The SK relation, in its most modern \citep{krumholz:2012apj} formulation, links the SFR ($\dot{\Sigma}_{\star}$) and total gas ($\Sigma$) surface density per unit free-fall time, $\dot{\Sigma}_{\star} = \epsilon_{\star}^{\rm ff} \Sigma/t_{\rm ff}$. The proportionality constant, often referred to as the efficiency per free-fall time following eq. \ref{eq_sk_relation}, is simply $\epsilon_{\star}^{\rm ff} = \zeta_{\rm sf} \fh2$. Experimentally, \citet{krumholz:2012apj} find $\epsilon_{\star}^{\rm ff} = 0.015$ (see \citealt{krumholz:2015review} for a complete review on the subject). This result is supported also by a larger set of observations including single MCs \citep{heiderman:2010,lada:2010apj}, local unresolved galaxies \citep{kennicutt:1998apj}, and moderate redshift, unresolved galaxies \citep{bouche2007apj,daddi:2010apj,daddi:2010b,tacconi:2010Natur,genzel:2010mnras}. The SK relation is shown in Fig.~\ref{fig_riassunto_sk}, along with the location of Dahlia and Alth{\ae}a at $z=6$.

Dahlia appears to be over-forming stars with respect to its gas mass, and therefore it is located about $3\sigma\,$ above the KS relation. As Alth{\ae}a needs about 10 times higher density to sustain the same SFR, its location is closer to expectations from the SK. We have checked that the agreement is even better if we use only data relative to MC complexes \citep[e.g.][]{heiderman:2010,murray:2011apj}. 

Dahlia's $\epsilon_{\star}^{\rm ff} = \zeta_{\rm sf} \fh2$ is similar to the analog values found by \citet{semenov:2015}, who compute such efficiency using a turbulent eddy approach \citep{padoan:2012}, with no notion of molecular hydrogen fraction. The difference is that Dahlia misses the high density gas. Alth{\ae}a instead matches both the $\epsilon_{\star}^{ff}$ and the amount of high density gas found by \citet{semenov:2015}. Also, its $\dot{\Sigma}_{\star}-\Sigma$ relation is consistent with \citet{torrey:2016arxiv}, who use a star formation recipe involving self-gravitating gas with a local SK \HH~dependent relation. 

From our simulations it is also possible to perform a cell-by-cell analysis of the SK relation (Fig.~\ref{fig_eos_sk}). As expected, the results show the presence of a consistent spread in the local efficiency values which, however, has a different origin for Dahlia and Alth{\ae}a. While in the former the variation is mostly due to a different enrichment level affecting \HH~abundance (eq. \ref{eq_fh2_anal}), for Alth{\ae}a the spread is larger because it results also from the individual evolutionary histories of the cells.

As noted by \citet{rosdahl:2017mnras}, for galaxy simulations with a SF model based on SK-like relation (eq. \ref{eq_sk_relation}), the resulting $\epsilon_{\star} =\epsilon_{\star}^{\rm ff}/t_{\rm ff}$ depends on how the feedback is implemented.
However, here we show that Alth{\ae}a has a lower $\epsilon_{\star}$ in spite of the fact that it implements exactly the same feedback prescription as Dahlia. The latter is qualitatively similar to a delayed cooling scheme used by \citet{rosdahl:2017mnras} and others \citep{stinson:2006mnras,teyssier:2013mnras}.
The lower efficiency $\epsilon_{\star}$ is a consequence of chemistry. As under non-equilibrium conditions the gas must be denser to form \HH, the ISM becomes more clumpy (Fig.~\ref{fig_morfologia}). These clumps can form massive clusters of OB stars which, acting coherently, yield stronger feedback and may disrupt completely the star forming site.

\subsection{Far and mid infrared emission}

A meaningful way to compare the two galaxies is to predict their \CIIion~and \HH~line emission, that can be observable at \highz with ALMA, and possibly with SPICA \citep[][in preparation]{spinoglio:2017,egami:2017}, respectively. Similarly to \P17cit, we use a modified version of the \CII~emission model from \citet[][hereafter \V15cit]{vallini:2015}. Such model is based on temperature, density and metallicity grids built using \textlcsc{cloudy} \citep{cloudy:2013}, as detailed in App.~\ref{sez_cloudy_model}. 
In Fig.~\ref{fig_mappe_comparison_2} we plot the \CII~$157.74\,\mu{\rm m}$ and \HH~$17.03\,\mu{\rm m}$ surface brightness maps ($S/(\surfl)$); the field of view is the same as in Fig.~\ref{fig_mappe_comparison_1}.

\subsubsection{Far infrared emission}

Let us analyze first the \CIIion~emission. Dahlia has a \CII~luminosity of $\log (L_{\rm CII}/\lsun) \simeq 7.5$ which is about 7 times smaller than Alth{\ae}a, i.e. $\log (L_{\rm CII}/\lsun) \simeq 8.3$. Fig.~\ref{fig_mappe_comparison_2} shows that the surface brightness morphology in the two galaxies is similar. Dahlia's emission is concentrated in the disk, featuring and average surface brightness of $\log\langle S_{\rm [CII]}/(\surfl) \rangle\simeq 6.4$ with peaks up to $\log(S_{\rm [CII]}/(\surfl))\simeq 7.4$ along the spiral arms. The analogous values for Alth{\ae}a are $7.3$ and $8.7$, respectively.

This can be explained as follows. 
FIR emission from the warm ($\simeq 10^{4} \rm K$), low density ($\lsim 0.1\,\cc$) component of the ISM is suppressed at \highz by the CMB (\citealt[][]{Gong:2012ApJ,dacunha:2013apj,pallottini:2015cmb}; \V15cit; App. \ref{sez_cloudy_model}), as the upper levels of the \CII~transition cannot be efficiently populated through collisions and the spin temperature of the transition approaches the CMB one \citep[see][for possibility of \CII~detection from low density gas via CMB distortions]{pallottini:2015cmb}.
Thus, $\simeq 95\%$ of the \CII~emission comes from dense ($\gsim 10\,\cc$, cold ($\simeq 100\,\rm K$), mostly molecular disk gas.
As noted in \V15cit (see in particular their Fig. 4) even when the CMB effect is neglected, the diffuse gas ($\lsim 0.1\,\cc$) account only for $\lsim 5\%$ of the emission for galaxies with ${\rm SFR}\sim 100 \msunyr$ and $Z\sim \zsun$, while it can be important in smaller objects \citep[][]{olsen:2015apj}.
The emissivity (in $\lsun/\msun$) of such gas can be written as in \P17cit \citep[in eq. 8, see also][]{Vallini:2013MNRAS,vallini:2017,goicoechea:2015apj}:
\be\label{eq_emission_cii}
\epsilon_{[\rm CII]} \simeq 0.1\, \left({n\over 10^{2} \cc}\right)\left({Z\over \zsun}\right)\,.
\ee
for $n\lsim 10^3\cc$, i.e. the critical density for \CII~emission\footnote{As the suppression of the CMB affects only the diffuse component ($\lsim 0.1\,\cc$), no significant difference is expected in the emissivity from the disks of the two galaxies (eq. \ref{eq_emission_cii}), that is composed of much higher ($\gsim 20\,\cc$) density material.}.
As the metallicity in the disk of the two galaxies is roughly similar ($\langle Z\rangle\simeq0.5\,\zsun$, see Tab.~\ref{tab_summary}), difference in the luminosities is entirely explained by the larger density in Alth{\ae}a. 
We stress once again that such density variation is a result of a more precise, non-equilibrium chemical network requiring to reach much higher densities before the gas is converted to stars. It is precisely that dense gas that accounts for a larger FIR line emissivity from PDRs.

We can also compare the calculated synthetic \CII~emission vs. SFR with observations (Fig.~\ref{fig_cii_sfr}) obtained for dwarf galaxies \citep{delooze:2014aa}, and available \highz detections or upper-limits. The \CII~emission from Dahlia is lower than expected based on the local \CII-SFR relation; its luminosity is also well below all upper limits for \highz galaxies. Although Alth{\ae}a is $\simeq 10 $ times more luminous, even this object lies below the local relation, albeit only by $1.3\sigma$. We believe that the reduced luminosity is caused by the combined effects of the CMB suppression and relatively lower $Z$. Note, however, that the predicted luminosity exceeds the upper limits derived for LAEs \citep[e.g.][]{ouchi2013,ota:2014apj}, but is broadly consistent with that of the handful of LBGs so far detected, like e.g. the four galaxies in the \citet{pentericci:2016apj}.
In general, observations are still rather sparse, with few \CII~detections with SFR comparable to Alth{\ae}a \citep[e.g.][]{capak:2015arxiv}. Also unclear is the amplitude of the scattering of the relation for \highz objects compared with local ones.
Improvements in the understanding of the ISM structure are expected from deeper observations and/or other ions \citep[e.g. \OIII][]{inoue:2016sci,carniani:2017bdf3299}. Also helpful would be a larger catalogue of simulated galaxies \citep[cfr.][]{ceverino:2017}, to control environmental effects.

\subsubsection{Mid infrared emission}

By inspecting the lower panel of Fig.~\ref{fig_mappe_comparison_2} showing the predicted \HH~$17.03\,\mu{\rm m}$~line emission, we come to conclusions similar to those for the \CII. Alth{\ae}a outshines Dahlia by $\simeq 15 \times$ by delivering a total line luminosity of $\log (L_{\rm H2}/\lsun) \simeq 6.5$. Differently from the \CII~case, also the deviations from the mean are much more marked in Alth{\ae}a, as appreciated from the Figure.

Note that the \HH~$17.03\,\mu{\rm m}$~line emissivity is enhanced in high density, high temperature regions. Indeed, \HH~emission mostly arises from shocked-heated molecular gas, for which $100\lsim n/\cc\lsim 10^5$ and for $10 \lsim T/{\rm K}\lsim 3000$ (see App.~\ref{sez_cloudy_model}).

For Dahlia, the disk density is relatively low, $n\simeq 30\,\cc$; in addition only $20\%$ of the gas is warm enough to allow some ($\epsilon_{\rm H2}\simeq 0.01\,\lsun/\msun$) emission. In practice, such emission predominantly occurs along the outer spiral arms of the galaxy where these conditions are met due to the heating produced by SN explosions. In the denser Alth{\ae}a disk, the gas emissivity can attain $\epsilon_{\rm H2}\simeq 10^{-3} - 0.01$ already at moderate $T = 200 \,\rm K$. The brightness peaks are associated to a few ($\lsim 1\%$) pockets of thousand-degree gas; they can be clearly identified in the map. 
This is particularly interesting because a galaxy like Alth{\ae}a might be detectable at very \highz with SPICA, as suggested by \citet[][in preparation]{egami:2017}.

\section{Conclusions}\label{sec_conclusione}
To improve our understanding of \highz galaxies we have studied the impact of \HH~chemistry on their evolution, morphology and observed properties. To this end, we compare two zoom-in galaxy simulations implementing different chemical modelling. Both simulations start from the cosmological same initial conditions, and follow the evolution of a prototypical $M_{\star}\simeq 10^{10}\msun$ galaxy at $z=6$ resolved at the scale of giant molecular clouds (30 pc). Stars are formed according to a \HH~dependent Schmidt-Kennicutt relation. We also account for winds from massive stars, SN explosions and radiation pressure in a stellar age/metallicity dependent fashion (see Sec.~\ref{sec_common_pre}). The first galaxy is named Dahlia and \HH~formation is computed from the \citet{krumholz:2009apj} equilibrium model; {Alth{\ae}a} instead implements a non-equilibrium chemistry network, following \citet{bovino:2016aa}. The key results can be summarized as follows:

\begin{itemize}

\item[\bf (a)] The star formation rate of the two galaxies is similar, and increases with time reaching values close to $100\, \msunyr$ at $z=6$ (see Fig.~\ref{fig_sfr_comparison}). However, Dahlia forms stars at a rate that is on average $1.5\pm 0.6$ times larger than Alth{\ae}a; it also shows a less prominent burst structure.

\item[\bf (b)] Both galaxies at $z=6$ have a SFR-stellar mass relation compatible with \citet[][]{jiang:2016apj} observations (Fig.~\ref{fig_sfr_mass_obs_comparison}). Moreover, they both show a continuous time evolution from specific SFR of ${\rm sSFR}\simeq 40\,{\rm Gyr}^{-1}$ to $5\,{\rm Gyr}^{-1}$. This is understood as an effect of the progressively increasing impact of stellar feedback hindering subsequent star formation events.

\item[\bf (c)] The non-equilibrium chemical model implemented in Alth{\ae}a determines the atomic to molecular hydrogen transition to occur at densities exceeding 300 $\cc$, i.e. about 10 times larges that predicted by equilibrium model used for Dahlia (Fig.~\ref{fig_chimica_krome}). As a result, Alth{\ae}a features a more clumpy and fragmented morphology (Fig.~\ref{fig_morfologia}). This configuration makes SN feedback more effective, as noted in point {(a)} above (Fig.~\ref{fig_energy_comparison}).

\item[\bf (d)] Because of the lower density and weaker feedback, Dahlia sits $3\sigma$ away from the Schmidt-Kennicutt relation; Alth{\ae}a, instead nicely agrees with observations (Fig.~\ref{fig_riassunto_sk}). Note that although the SF efficiency is similar in the two galaxies and consistent with other simulations \citep{semenov:2015}, Dahlia is off the relation because of insufficient molecular gas content (Fig.~\ref{fig_eos_h2}).

\item[\bf (e)] We confirm that most of the emission from the \CIIion~and \HH~is due to the dense gas forming the disk of the two galaxies. Because of Dahlia's lower average density, Alth{\ae}a outshines Dahlia by a factor of $7$ ($15$) in \CII~$157.74\,\mu{\rm m}$ (\HH~$17.03\,\mu{\rm m}$) line emission (Fig.~\ref{fig_mappe_comparison_2}). Yet, Alth{\ae}a has a 10 times lower \CII~luminosity than expected from the locally observed \CII-SFR relation (Fig.~\ref{fig_cii_sfr}). Whether this relation does not apply at \highz or the line luminosity is reduced by CMB and metallicity effects remains as an open questions which can be investigated with future deeper observations. 

\end{itemize}

To conclude, both Dahlia and Alth{\ae}a follow the observed \highz SFR-$M_{\star}$ relation. However, many other observed properties (Schmidt-Kennicutt relation, \CIIion~and \HH~emission) are very different. This shows the importance of accurate, non-equilibrium implementation of chemical networks in early galaxy numerical studies.

\section*{Acknowledgments}
We are grateful to the participants of \emph{The Cold Universe} program held in 2016 at the KITP, UCSB, for discussions during the workshop.
We thank P. Capelo, D. Celoria, E. Egami, D. Galli, T. Grassi, L. Mayer, S. Riolo, J. Wise for interesting and stimulating discussions.
We thank the authors and the community of \textlcsc{ramses} and \textlcsc{pymses} for their work.
AP acknowledges support from Centro Fermi via the project CORTES, \quotes{Cosmological Radiative Transfer in Early Structures}.
AF acknowledges support from the ERC Advanced Grant INTERSTELLAR H2020/740120.
RM acknowledge support from the ERC Advanced Grant 695671 \quotes{QUENCH} and from the Science and Technology Facilities Council (STFC).
SS acknowledges support from the European Commission through a Marie Sk{\l}odowska-Curie Fellowship, program PRIMORDIAL, Grant No. 700907.
This research was supported in part by the National Science Foundation under Grant No. NSF PHY11-25915.

\bibliographystyle{mnras}
\bibliography{master}
\bsp

\appendix
\section{Minkowsky functionals}\label{sec_app_minchioschi}

In general, Minkowsky functionals are mathematical tools that give a complete characterization of the morphology of a ${\rm I\!R}^{n}\mapsto {\rm I\!R}$ field. In astrophysics they have been proposed as a mean to give a description of the large scale structure \citep[e.g.][]{Mecke:1994}, to study the topology of \HII~bubbles for reionization studies \citep[e.g.][]{gleser:2006mnras,yoshiura:2017mnras}, and (for $n=2$ fields) analyze CMB anisotropies and non-gaussianity \citep[e.g.][]{schmalzing:1998,novikov:2000aa}.

A formal definition can be given following \citet[][]{schmalzing:1997apj}. Let $u(\mathbf{x})$ denote a scalar field defined on a subset of ${\rm I\!R}^{3}$ with volume $V$. Let us take $u$ such that it has zero mean ($\langle u(\mathbf{x})\rangle = 0$) and variance $\langle u^2(\mathbf{x})\rangle = \sigma$. Then we can define the excursion set $F_{\nu}(\mathbf{x})$ as the ensemble of regions in $V$ satisfying $u(\mathbf{x})>\nu \sigma$. Then, the Minkowsky functionals can be defined in terms of volume (${\rm d}^3 x$) and surface (${\rm d}^2 x$) integrals as a function of the threshold $\nu$:
\begin{subequations}\label{eq_mink_definition}
\begin{align}
V_{0}(\nu) =&  (V     )^{-1} \int_{V}               \Theta(u -\nu\sigma) {\rm d}^3 x\\
V_{1}(\nu) =&  (6V    )^{-1} \int_{\partial F_{\nu}}                     {\rm d}^2 x\\
V_{2}(\nu) =&  (6\pi V)^{-1} \int_{\partial F_{\nu}} (\kappa_1 +\kappa_2){\rm d}^2 x\\
V_{3}(\nu) =&  (4\pi V)^{-1} \int_{\partial F_{\nu}} \kappa_1 \kappa_2   {\rm d}^2 x,\,
\end{align}
\end{subequations}
where $\Theta$ is the Heaviside function, $\partial F_{\nu}(\mathbf{x})$ the surface beading the excursion set $F_{\nu}$, and $\kappa_{1}(\mathbf{x})$ and $\kappa_{2}(\mathbf{x})$ the two principal curvatures of the surface. In practical terms, $V_{0}$ is a measure of the volume filling factor of the excursion set with threshold $\nu$, $V_{1}$ of the surface, $V_{2}$ of the mean curvature (sphericity/concavity) and $V_{3}$ of the Euler characteristic (shape of components).

The curvatures on the surface $\partial F_{\nu}$ can be expressed via the Koenderink invariant \citep[][see appendix A and reference therein]{gleser:2006mnras}: by adopting the Einstein sum convention we can write
\begin{subequations}
\begin{align}
\kappa_1 + \kappa_2 =& \epsilon^{ijk} \epsilon^{lmn} \delta_{kn} (\partial_i u) (\partial_j\partial_l u) (\partial_m u) / N_{t}^{3/2}\\
2\kappa_1 \kappa_2  =& \epsilon^{ijk} \epsilon^{lmn} (\partial_i u) (\partial_l u) (\partial_j \partial_m u) (\partial_k \partial_n u) / N_{t}^{2}\\
N_{t} =& (\partial_p u) (\partial^p u)\,,
\end{align}
\end{subequations}
where $\epsilon_{ijk}$ is the Levi-Civita symbol, $\delta_{ij}$ is the Kronecker delta, and $\partial_i$ is the $i$-th component of the partial derivative operator. Finally, eq.s \ref{eq_mink_definition} can be suitably expressed as integral over the volume by using the following relation
\be
\int_{\partial F_{\nu}} {\rm d}^2 x = \int_{V} \delta(u-\nu\sigma) N_{t}^{1/2} {\rm d}^3 x\,,
\ee
where $\delta$ is the Dirac delta.

As an illustrative example, we can compute the Minkowsky functionals for a zero-mean Gaussian random field $u\equiv\log\Delta$ with variance $\sigma^2 = \langle\log\Delta^2\rangle$ and variance of the tangent field $\sigma_{t}^2 = \langle (\partial_p \log\Delta)(\partial^p \log\Delta) \rangle$. For such Gaussian field, the Minkowsky functionals can be expressed using the following analytical expression \citep[][see also \citealt{gleser:2006mnras}]{schmalzing:1997apj}
\begin{subequations}
\begin{align}\label{eq_minkowsky_gauss_anal}
V_{0} =& 1/2 - c_0 \int_{0}^{\nu} \exp(-x^2 / 2) {\rm d}x \\
V_{1} =& c_1 \lambda \exp(-\nu^2 / 2)\\
V_{2} =& c_2 \lambda^2 \nu \exp(-\nu^2 / 2)\\
V_{3} =& c_3 \lambda^3 (\nu^2 -1) \exp(-\nu^2 / 2)\,,
\end{align}
\end{subequations}
where $\nu = \log\Delta / \sigma$, $\lambda = (6\pi)^{-1/2}\sigma_{t}/\sigma$, and $c_{i}$ are numerical constant with values $c_0 = c_3 = (2\pi)^{-1/2}$ and $c_1 = c_2 = 2/(3(2\pi)^{1/2})$.

\begin{figure}
\centering
\includegraphics[width=0.49\textwidth]{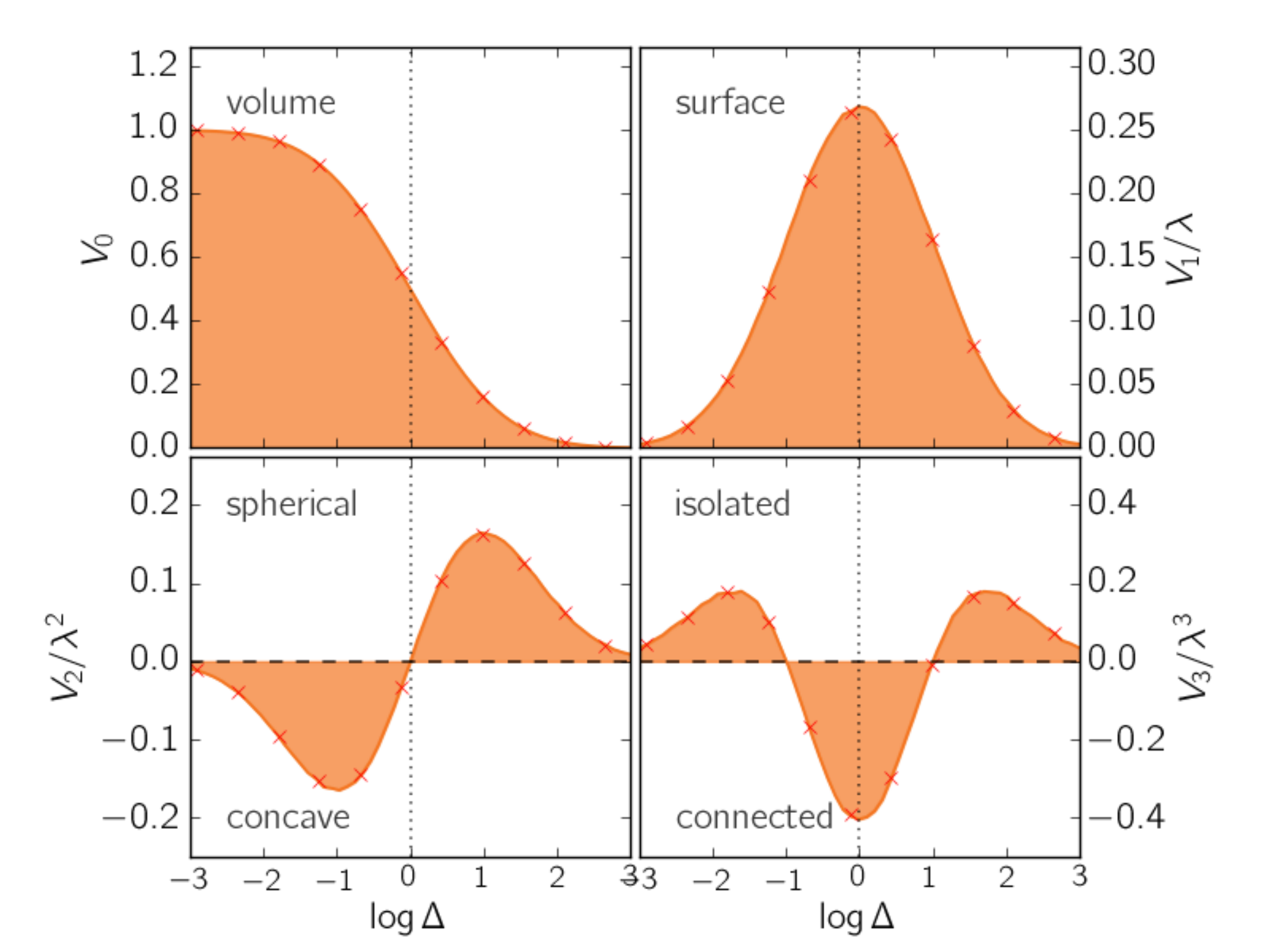}
\caption{
Example of Minkowsky functionals calculated for the zero-mean Gaussian random field $\log\Delta$. The functionals are plotted with a with a orange line and transparent region and are normalized by $V_{i}/\lambda^{i}$, with $\lambda = (6\pi)^{-1/2}\sigma_{t}/\sigma$, where $\sigma =1$ is the field variance and $\sigma_{t}\simeq 5.6 /{\rm kpc}$ is the variance of the tangent field. Analytical expected values for the functionals (eq.s \ref{eq_minkowsky_gauss_anal}) are plotted with red crosses. To guide the eye, $\log \Delta = 0$ is marked with a dotted vertical black line, and both $V_{2} = 0$ and $V_{3} = 0$ are highlighted with dashed horizontal lines.
\label{fig_morfologia_test}
}
\end{figure}

We numerically compute the Minkowsky functionals for th e $\log\Delta$ field  with $\sigma =1$ on a $256^3$ unigrid box with volume $(10\,{\rm Mpc})^{3}$, that thus result in a tangent field variance of $\sigma_{t}\simeq 5.6 /{\rm kpc}$. The resulting Minkowsky functionals are plotted in Fig.~\ref{fig_morfologia_test}. We find a very good match with the analytical values.

Since the chosen $\log\Delta$ Gaussian field is an approximation to the quasi-linear regime of the cosmic density field \citep[e.g.][]{coles:1991mnras,choudhury:2001mnras}, it is intuitive to analyze the properties of its Minkowsky functionals. In Fig.~\ref{fig_morfologia_test} the filling factor $V_{0}$ gives the probability of finding regions with increasing overdensity $\Delta$; note that at $\Delta =1$ (mean density), $V_{0} = 0.5$, i.e. it is equiprobable to find voids ($\log\Delta\lsim -1$) and overdense regions ($\log\Delta\gsim 1$). Both voids and overdense regions are isolated $V_{3}(\gsim 1)=V_{3}(\lsim -1) > 0$ and have a smaller area with respect to mean density regions ($V_{1}(\gsim 1)=V_{1}(\lsim-1) \lsim V_{1}(0)$). However, while overdensities have spherical shapes $V_{2}(\gsim 1)>0$, voids are concave regions ($V_{2}(\gsim 1)<0$): both voids and overdense regions are delimited by connected ($V_{2}(\simeq 0)<0 $) mean density regions, that are almost flat ($V_{2}(\simeq 0)\simeq 0$) and have very large surface areas.

\section{Emission from \CIIion~and \HH.}\label{sez_cloudy_model}

To compute the emission from \CIIion~ions and \HH~molecules, we post-process the simulation outputs using the photoionization code \textlcsc{cloudy} \citep{cloudy:2013}, similarly to what done in \V15cit~and \P17cit. We consider a grid of models based on the density ($n$), temperature ($T$) and metallicity ($Z$) of the gas in our simulation. We produce a total of $10^3$ models, that are parameterized as a function of the column density ($N$). For each model we adopt a plane-parallel geometry and assume a dust content proportional to the metallicity.

The radiation field includes the CMB background and an interstellar radiation field produced by stars, that is obtained by rescaling the Milky Way spectrum \citep{black:1987} using the main galaxy (Dahlia or Alth{\ae}a) SFR. At $z=6$ Dahlia has a star formation rate ${\rm SFR} = 156\, \msunyr$ and Alth{\ae}a has ${\rm SFR} = 136\,\msunyr$, where the uncertainty is the r.m.s. in the last $50\,\myr$ (Sec.~\ref{sec_sfr_feed}, in particular see Fig.~\ref{fig_sfr_comparison}). For modelling convenience, in the \textlcsc{cloudy} calculation we set ${\rm SFR} = 100\,\msunyr$. Note that a larger value for the rescaling does not yield a large variation of the expected \CII~in molecular gas \citep[][ with $G=G_{0}\,{\rm SFR}/\msunyr$]{vallini:2017}, and \HH~emission is relatively unaffected by the field, as the excitation is mostly due to shocks \citep[e.g.][]{black:1976apj,ciardi:2001}.

As noted in Sec.~\ref{sec_chem_noneq}, accounting for the UVB is not relevant for the ionization state of the gas in the proximity of galaxies \citep[][]{gnedin:2010}. Thus, in our \textlcsc{cloudy} models we consider that the gas is shielded by a column density of $N\simeq 10^{20}{\rm cm}^{-2}$.

Regarding the \CII~we underline that the effect of CMB suppression of \CII~is included in the \V15cit~model \citep[see also]{dacunha:2013apj,pallottini:2015cmb}. Such effect suppress the emission where the spin temperature of the \CII~transition is close to the CMB one. This is relevant for low density ($n\lsim 10^{-1}\,\cc$) medium, that does not have enough collisions to decouple from the CMB. Here we do not account for the photoevaporation effect on MC, that has an important impact on FIR emission \citep[][]{vallini:2017}, particularly when including a spatially varying FUV field, not included in the present modelling.

Even though in the present paper we only show the \HH~line at $\lambda = 17.04\, \mu{\rm m}$ (Sec.~\ref{sec_obs_prop}, in particular see Fig.~\ref{fig_mappe_comparison_2}), we used \textlcsc{cloudy} to compute the emission of the following \HH~roto-vibrational lines: 0-0 S(0), 0-0 S(1), 0-0 S(5), and 1-0 S(1), that correspond to transition of wavelength $\lambda/\mu{\rm m} = 2.12,\, 6.91,\, 9.66,\, 17.04,\,{\rm and}\, 28.22$, respectively \citep[see][in preparation]{spinoglio:2017,egami:2017}.

For the considered five transitions, the oscillator strength is a decreasing function of $\lambda$ \citep[e.g.][]{turner1977apjs}, going from $\simeq 3\,\times 10^{-7}\,{\rm s}^{-1}$ for $\lambda = 2.12\,\mu{\rm m}$ to $\simeq 3\,\times 10^{-11}\,{\rm s}^{-1}$ for $\lambda = 28.22\,\mu{\rm m}$. On the other hand, both the excitation temperature \citep{timmermann:1996} and the critical density for collisional excitation \citep{roussel:2007apj} decrease for decreasing $\lambda$ \citep[see also][]{black:1976apj}, e.g. for $\lambda = 2.12\,\mu{\rm m}$ we have an excitation temperature $T_{\rm ex}\simeq 6\times10^4{\rm K}$ and a critical density $n_{cr} \simeq 10^4 \cc$, while for $\lambda = 28.22\,\mu{\rm m}$ we have $T_{\rm ex}\simeq 5\times10^3{\rm K}$ and $n_{cr} \simeq 5\,\cc$. In the simulations the bulk of the \HH~gas have $\langle T\rangle \sim10^2{\rm K}$ and density in the range $10^2 \lsim n/\cc\lsim 10^3$ (see Fig.~\ref{fig_eos_h2}); thus in both Dahlia and Alth{\ae}a the most favoured \HH~transition is the $\lambda = 17.03\mu{\rm m}$, followed by the $28.21\mu{\rm m}$ line.

\label{lastpage}

\end{document}